\documentclass{iopjournal}
\usepackage{amsmath, amssymb, mathtools, mathrsfs}
\usepackage[colorlinks=true]{hyperref}
\hypersetup{colorlinks,linkcolor={red},citecolor={blue},urlcolor={blue}}  
\usepackage{color}
\usepackage{graphicx}
\usepackage[normalem]{ulem}
\usepackage{mathtools}
\usepackage{stmaryrd}
\usepackage{float}
\usepackage{textalpha}
\usepackage{siunitx}
\usepackage{comment}
\usepackage{soul}

\usepackage{cite}
\usepackage{ragged2e}

\newcommand{\bs}[1]{{\boldsymbol{#1}}}
\newcommand{\bk}{\bs{k}}
\newcommand{\br}{\bs{r}}

\newcommand{\bq}{\bs{q}}
\newcommand{\bp}{\bs{p}}

\newcommand{\Ef}{E_\mathrm{f}}
\newcommand{\Ec}{E_\mathrm{c}}
\newcommand{\VR}{V_\mathrm{R}}

\newcommand{\fc}{f_\mathrm{c}}

\newcommand{\Ket}[1]{\mathop{ | #1 \rangle}}
\newcommand{\Bra}[1]{\mathop{ \langle #1 |}}
\newcommand{\BraKet}[2]{\mathop{ \langle #1 | #2 \rangle}}
\newcommand{\dd}{\mathrm{d}}

\newcommand{\LCF}{Universit\'e Paris-Saclay, Institut d'Optique Graduate School, CNRS, Laboratoire Charles Fabry, 91127, Palaiseau, France}
\newcommand{\LKB}{Laboratoire Kastler Brossel, Sorbonne Universit\'e, CNRS, ENS-PSL Research University, Coll\`ege de France, 4 Place Jussieu, 75005 Paris, France}
\newcommand{\STE}{Universit\'e Jean Monnet Saint-Etienne, CNRS, Institut d'Optique Graduate School,
Laboratoire Hubert Curien, UMR 5516, Saint-Etienne F-42023, France }

\begin{document}

\articletype{Paper} %	 e.g. Paper, Letter, Topical Review...

\title{Energy-resolved transport of ultracold atoms across the Anderson transition: theory and experiment}

\author{Jean-Philippe Banon$^{1,2,3,*}$, Sacha Barr\'e$^1$, Ke Xie$^1$, Hoa Mai Quach$^1$, Xudong Yu$^1$, Yukun Guo$^1$, Myneni Niranjan$^1$, Alain Aspect$^1$, Vincent Josse$^1$, Nicolas Cherroret$^{2}$ }

\affil{$^1$ \LCF}

\affil{$^2$ \LKB}

\affil{$^3$ \STE}

\affil{$^*$Author to whom any correspondence should be addressed.}

%\email{nicolas.cherroret@lkb.upmc.fr}
\email{jean.philippe.banon@univ-st-etienne.fr}

\justifying

\begin{abstract}
\vspace{0.3cm}
\small
\justifying
In a recent experiment [X. Yu et al., arXiv:2602.07654], energy-resolved measurements of an atomic matter wave spreading in a speckle potential enabled the direct observation of the three-dimensional Anderson transition. In this work, we present a quantitative theoretical description of the matter-wave dynamics based on a tailored implementation of the self-consistent theory of localization, which incorporates both the spectral and spatial properties of the state prepared in the disorder.
%This approach provides a general theoretical framework well suited for the quantitative analysis of wave-packet dynamics in three-dimensional disordered systems. 
We benchmark this theoretical approach against \emph{ab initio} numerical simulations, and use it to analyze the atom density profiles observed experimentally in the localized, diffusive, and critical regimes. Particular emphasis is placed on the key role of the atomic energy distribution, especially on the distinct contributions of Bose-condensed and thermal atoms to interpret the experimental profiles. 
Our framework provides a versatile and  efficient theoretical toolbox for quantitatively describing wave-packet dynamics in three-dimensional disordered quantum systems, which remains challenging for state-of-the-art large-scale numerical simulations.
\end{abstract}

\section{Introduction}
\vspace{0.3cm}

Understanding quantum transport in disordered media is a long-standing challenge, central to condensed-matter \cite{Lee1985}, wave physics \cite{Lagendijk2009}, and cold-atom experiments \cite{Aspect2009}. A key phenomenon in this context is Anderson localization, whereby interference effects suppress diffusion and can ultimately lead to the absence of transport \cite{Anderson1958, Abrahams1979, Abrahams2010}. Ultracold atoms provide a particularly powerful platform to investigate this physics, as disorder, dimensionality, and interactions can be controlled with high precision, while observables such as density profiles are directly accessible \cite{Clement2006, sanchez2010disordered, Mueller2011, Shapiro2012, Cherroret2021}. 
These capabilities have stimulated an intense theoretical activity \cite{Sanchez-Palencia2007, Skipetrov2008, Lugan2009, Filoche2012, Piraud_2012, CFS_Karpiuk2012, Plisson2013, Piraud2014, delande2014mobility, Fratini2015, muller2016critical, Pasek2017, Lemarie2019, filoche2024, Vrech2026}, and enabled experimental demonstrations of Anderson localization in one-dimensional \cite{Billy2008, Roati2008}, two-dimensional \cite{White2020} and three-dimensional (3D)~\cite{kondov2011three, Jendrzejewski2012,Semeghini2015,Barbosa2024} disordered potentials, as well as precise studies of analogous dynamical localization phenomena in kicked-rotor systems \cite{Moore1995, Chabe2008, Hainaut2017, Hainaut2018, Arrouas2026, Madani2025}.

In a recent experiment \cite{Josse2026}, atoms were transferred using a radio-frequency (rf) scheme into a 3D disordered potential with a well-controlled and narrow energy distribution~\cite{Volchkov2018,Lecoutre2022}. This protocol enables the exploration of wave-packet dynamics at selected energies across the localized, diffusive, and critical regimes of the Anderson transition. In particular, it overcomes difficulties encountered in previous setups~\cite{kondov2011three,Jendrzejewski2012,Semeghini2015, Barbosa2024}, which were limited by a broad  atomic energy distribution that prevented the direct observation of the transition and the study of the critical regime (see, e.g., Ref.~\cite{Pasek2017}).

While the measurements reported in Ref.~\cite{Josse2026} provided a direct observation of the Anderson transition, a fully quantitative connection to microscopic theory remains to be established, in particular for an accurate description of the experimentally observed spatial density profiles. This calls for the development of a dedicated theoretical framework capable of reliably describing wave propagation over large distances and long time scales across the mobility edge in three dimensions. Such a framework must incorporate the key experimental ingredients, notably the spatial and spectral properties of the initial states prepared by the rf transfer in the disorder. 

In the present work, we undertake this task by developing a tailored implementation of the 3D self-consistent theory (SCT) of localization \cite{VW1980a, VW1980b, Vollhardt10}, designed to capture the above features of wave-packet dynamics. We first benchmark this approach against \emph{ab initio} numerical simulations of 3D wave-packet spreading, which explicitly implement the rf-loading scheme and reproduce the statistical properties of the experimental speckle disorder. We then confront the theoretical predictions with experimental measurements of the density profiles across the mobility edge. This analysis reveals the central role of the atomic energy distribution in shaping the spatial structure of wave packets in the disordered potential.
%\textcolor{red}{in particular by clarifying the distinct contributions of atoms belonging to the condensate and to the thermal components of the cloud. VJ : really needed ? : could be simply mentioned in the outline below}
\newline

The article is organized as follows. In Sec.~\ref{Sec:theory}, we recall the main elements of the experiment reported in Ref.~\cite{Josse2026}, which enabled the realization of energy-resolved propagation of atomic wave packets in a disordered potential. We then develop a theoretical framework adapted to the experimental scenario, based on the SCT of localization. In Sec.~\ref{Sec:numerics}, we benchmark this approach against \emph{ab initio} numerical simulations performed under conditions close to those of the experiment. Section~\ref{Sec:experiments} is devoted to the comparison with the experimental measurements, including a characterization of the spatial and spectral distributions of the atomic cloud transferred into the disorder. In particular, we discuss the respective contributions of condensed and thermal atoms to the energy distribution.
Finally, we summarize our findings and outline perspectives in Sec.~\ref{Sec:conclusion}.

\section{Energy-resolved wave packet propagation across the Anderson transition: theory}
\label{Sec:theory}
\vspace{0.3cm}

\subsection{Experimental context}

In the experiment of Ref.~\cite{Josse2026}, the dynamical spreading of ultracold atomic clouds prepared with well-defined energies in a three-dimensional random speckle potential was investigated in the vicinity of the Anderson mobility edge. Such energy-resolved transport was achieved using the scheme illustrated in Fig.~\ref{Fig:scheme}: 
\begin{figure}[h]
\centering
\includegraphics[scale=0.6]{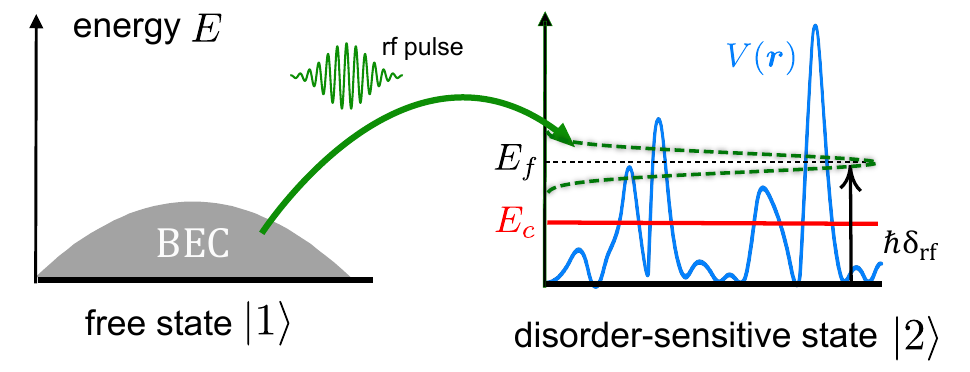}
\caption{ \justifying Experimental protocol of Ref.~\cite{Josse2026}: atoms from a Bose–Einstein condensate, initially prepared in a disorder-free state, are transferred to a disorder-sensitive state by means of an rf pulse. The transferred atoms acquire a narrow energy distribution (green dashed curve) centered at $\Ef$, which can be tuned across the mobility edge $\Ec$ of the disordered potential.}
\label{Fig:scheme}
\end{figure}
atoms from a Bose–Einstein condensate (BEC) are first prepared in a disorder-free hyperfine state $|1\rangle$, then transferred to a disorder-sensitive state $|2\rangle$ at a well-defined target energy $\Ef=\hbar\delta_\text{rf}$. The transfer is driven by a radio-frequency pulse whose frequency sets the detuning $\delta_\text{rf}$ from the bare resonance of state $|2\rangle$. As shown in \cite{Josse2026,Volchkov2018}, in the weak-coupling regime where only a small fraction of the BEC is transferred, and provided that the condensate is large enough to be approximated by the uniform plane-wave state $|\bk=0\rangle$, the atoms loaded into the disorder acquire the following energy distribution:
\begin{align}
\label{eq:energy_dist}
\mathcal{D}(E;\Ef)\simeq F(E-\Ef) A(E,\bk=0) \, .
\end{align}
Here $A(E,\bk=0)$ is the zero-momentum spectral function in the disordered potential, i.e., the probability density that an initial plane-wave state $|\bk=0\rangle$  has energy $E$ in the random potential \cite{Volchkov2018}, and $F(E-\Ef)$ is an energy-filter function determined by the temporal shape of the rf pulse. The filter is peaked around the target energy $\Ef$, with a typical width $\Delta E \ll \Ef$. Note that the spectral function alone would correspond to the energy distribution obtained after an abrupt switch-on of the disorder applied to the BEC (see, e.g., Ref.~\cite{Jendrzejewski2012}). This protocol thus enables the preparation of atomic wave packets with a much narrower energy distribution than in previous experiments. Moreover, the target energy $\Ef$ can be finely tuned across the mobility edge $\Ec$, allowing for a precise investigation of the Anderson transition.

\subsection{Energy-resolved state: model }
\vspace{0.3cm}

Having introduced the experimental protocol, we now develop a simple model describing the preparation and the subsequent evolution of an energy-resolved state in the random potential. We denote by $|\phi\rangle$ the condensate wave function prior to the transfer, and by $|\psi_0\rangle$ the state of the atoms loaded into the disorder. For a coherent transfer, as realized in \cite{Josse2026}, we model the relation between these two states by
\begin{equation}
    \Ket{\psi_0} = f(\hat{H}-\Ef)|\phi\rangle= \sum_{n} \Ket{\phi_n} \BraKet{\phi_n}{\phi} f(E_n-\Ef) \:,
    \label{eq:expansion_psi0}
\end{equation}
\sloppy where $\hat{H}=\bp^2/(2m) + V(\br)$ is the Hamiltonian in the disorder-sensitive state, with $V(\br)$ the speckle potential, and $f(E-\Ef)$ is an amplitude filter function such that $\smash{|f(E-\Ef)|^2\equiv F(E-\Ef)}$. In the second equality, we have introduced a complete set  $(|\phi_n\rangle)$ of eigenstates of $\hat{H}$ with energies $E_n$, which will prove convenient in the following. Experimentally, the filter function $f$ is directly linked to the Fourier transform of the temporal envelope of the rf pulse, as given by first order perturbation theory for an initial state transferred to a continuum. Equation~\eqref{eq:expansion_psi0} is therefore valid in the weak-coupling regime and provided interactions can be neglected. One can easily verify that the definition (\ref{eq:expansion_psi0}) correctly yields the energy distribution \eqref{eq:energy_dist} of the atoms in the disorder. Indeed, we have by definition
 \begin{align}
 \label{eq:DE_derivation}
   \mathcal{D}(E;\Ef)\equiv\overline{\Bra{\psi_0} \delta \big( E - \hat{H} \big) \Ket{\psi_0}}
   &\simeq
   |f(E-\Ef)|^2\sum_n \overline{\left|\BraKet{\phi_n}{\bk=0} \right|^2 \delta \left( E - E_n \right)}\nonumber\\
   & \equiv F(E-\Ef)A(E,\bk=0),
 \end{align}
where Eq.~\eqref{eq:expansion_psi0} has been used in the second equality, together with the approximation $|\phi\rangle\simeq|\bk=0\rangle$ for the condensate wave function in state $|1\rangle$. Here and in the following, the overline denotes averaging over disorder realizations.

% {\color{red}JP: Should we precise here the condition for which the approximation $\Ket{\phi} \simeq \Ket{\bk=0}$ is valid (in fact it is rather $A(E,\phi) \simeq A(E,\bk=0)$)? Later we show a rather narrow Gaussian wave function which may raise questions. Maybe we could also justify this by showing in Appendix C.4 that the energy distributions for $\Ket{\bk=0}$ and for $\Ket{\phi}$ are indeed close?}

%, the  state transferred in the disorder can be rewritten as
%\begin{equation}
%    \Ket{\psi_0} = \sum_{n} \Ket{\phi_n} \BraKet{\phi_n}%{\phi} f(E_n) \:.
%    \label{eq:expansion_psi0}
%\end{equation}
%In Ref.~\cite{Josse2026}, the filter function is given by the Fourier transform of a Kaiser time window function used for the radio-frequency coupling (see Sec.~\ref{Sec:experiments}). In this section, however, we keep the discussion general and compute the disorder-averaged density $n(\br,t)\equiv \overline{|\psi(\br,t)|^2}$ at a  time $t$ after the loading process and with an arbitrary shape for $f(E)$.

\subsection{Disorder-averaged density}
\label{Sec:average_densitytheo}
\vspace{0.3cm}

We now derive an expression for the disorder-averaged density $n(\br,t)$, suitable for the application of the self-consistent theory of localization, starting from the filtered initial state~\eqref{eq:expansion_psi0}. 
The wave function $\psi(\br,t)$ of the cloud in the random potential at time $t$ is related  to the initial state $\psi(\br,t=0) = \psi_0(\br)$ by
\begin{align}
    \psi(\br,t)=\Theta(t) \Bra{\br} \exp \big(\!-i\hat{H} t \big) \Ket{\psi_0} = \Theta(t)
    \sum_{n} \BraKet{\br}{\phi_n} \BraKet{\phi_n}{\phi} f(E_n-\Ef) \exp \left( -i E_n t \right) \: ,
\end{align}
where $\Theta$ is the Heaviside step function 
%and where Eq.~\eqref{eq:expansion_psi0} has been used in the second equality 
(unless otherwise stated, from now on we set $\hbar=1$). By inserting the closure relation $\mathbf{1} = \int \Ket{\br^\prime}\Bra{\br^\prime} \: \mathrm{d}\br^\prime$ and using the spectral representation
\begin{equation}
    \Theta(t) f(E_n-\Ef)\exp(-iE_n t) = -i \int \frac{\dd E}{2\pi}\frac{f(E-\Ef) \exp\left( -i E t\right)}{E-E_n+i0}  \: ,
\end{equation}
the wave function can be rewritten as
\begin{align}
    \psi(\br,t>0)= -i \int \frac{\dd E}{2\pi}
    \int \dd \br' G^{R}(\br,\br^\prime,E) \phi(\br^\prime) f(E-\Ef) \exp \left( -i E t \right)  \: ,
    \label{eq:psi_t_exact} 
\end{align}
where we have introduced the retarded Green's function
\begin{equation}
    G^{R}(\br,\br^\prime,E) = \sum_{n} \frac{\BraKet{\br}{\phi_n} \BraKet{\phi_n}{\br^\prime}}{E - E_n + i0}  \: .
\end{equation}
From Eq.~(\ref{eq:psi_t_exact}), we obtain a general expression for the disorder-averaged density:
\begin{align}
\label{eq:density_WP}
n(\br,t)=\!
 \int \dd \br' \dd \br''\!\int \frac{\dd E}{2\pi}\frac{\dd \omega}{2\pi}e^{-i\omega t} &\overline{G^R(\br,\br',E_+)G^{A}(\br'',\br,E_-)}
 \phi(\br')\phi^*(\br'')\nonumber\\
 &\times f(E_+-\Ef)f^*(E_--\Ef),
\end{align}
where  $G^{A}(\br'',\br,E)\equiv G^{R*}(\br,\br'',E)$ is the advanced Green’s function, and we have changed the energy variables to  $E\equiv (E_1+E_2)/2$ and $\omega\equiv E_1-E_2$, with $E_\pm=E\pm\omega/2$.
Although Eq.~(\ref{eq:density_WP}) is fully general, it is not convenient for practical use, as it involves a correlator of Green’s functions evaluated at three distinct spatial points. However, in the so-called hydrodynamic regime, where fast spatial and temporal variations of the profile  (typically occurring on the scale of the mean free path and the mean free time, respectively) are neglected, this expression can be significantly simplified. The derivation is presented in Appendix~\ref{App:theory} for clarity (see also Refs.~\cite{Cherroret2016, Shapiro2012}), and leads to
\begin{align}
 \label{eq:density_full}
n(\br,t)\simeq\int \dd E\int \dd \br' 
\mathcal{D}(E;\Ef)P_{E}(\br,\br',t)|\phi(\br')|^2.
\end{align}
In this relation, $\mathcal{D}(E;\Ef)$ is the energy distribution (\ref{eq:energy_dist}), and $P_E$ 
is the disorder-averaged propagator, defined as
\begin{equation}
\label{eq:propagator_def}
    P_E(\br,\br',t)\equiv \frac{1}{2\pi\rho}\int\frac{\dd \omega}{2\pi}
    e^{-i\omega t}\overline{G^R(\br,\br',E_+)G^{A}(\br',\br,E_-)},
\end{equation}
with $\rho$ the density of states per unit volume. Physically, $P_E$  gives the probability density for a particle with energy $E$ to propagate from $\br'$ at time $t=0$ to $\br$ at time $t$ in the random potential \cite{Akkermans07}. 
The averaged density (\ref{eq:density_full}) then follows by summing over all possible initial positions $\br'$ and all  energies, weighted by the energy distribution of the atoms. In the ideal limit of a very  sharp filter function $F(E-\Ef)\propto\delta(E-\Ef)$, Eq. (\ref{eq:density_full}) further simplifies to 
\begin{align}
 \label{eq:density_narrow}
n(\br,t)\simeq \int \dd \br'  P_{\Ef}(\br,\br',t)|\phi(\br')|^2,
\end{align}
so that the density becomes a direct probe of transport in the disorder at energy $\Ef$. \newline

In Sec.~\ref{Sec:numerics}, Eq.~(\ref{eq:density_narrow}) will be compared with numerical simulations performed in the ideal sharp-filter limit. By contrast, when comparing with the experimental data in Sec.~\ref{Sec:experiments}, it will be necessary to account for the finite width of the energy filter. We will therefore rely instead on Eq.~(\ref{eq:density_full}).%, extending the expression (\ref{eq:energy_dist}) of $\mathcal{D}(E;\Ef)$ to incorporate the contribution of thermal atoms.

%{\color{red}
%\medskip
%[JPB: Here we end up with the result that one should convolve $|\phi|^2$ with the propagator. I feel it will be in tension with what we do afterward, especially in the experiment where we strictly speaking use $\overline{|\psi_0|^2}$ instead of $|\phi|^2$. I agree that this feels somewhat more natural but this is a little bit in contradiction with the above equation. I think we should justify somewhere why this is fine and reasonable. I will write some ideas of justification about this small issue in a separate document.} 

\subsection{Propagator: self-consistent theory}
\vspace{0.3cm}
\label{Sec:Propagator}

We now derive a general expression for the propagator $P_E$ capturing the wave-packet dynamics across the mobility edge of the Anderson transition. To this end, we use the self-consistent theory of localization, a simple model in which the quantum interference responsible for Anderson localization are modeled by a re-summation of Cooper diagrams \cite{VW1980a, VW1980b, Vollhardt10, Cherroret2016}.  Within this approach,  the Fourier transform of the propagator (\ref{eq:propagator_def}) is given by
\begin{equation}
\label{eq:Pqomega}
P_E(\bq,\omega)=\frac{1}{-i\omega+D(\omega)q^2},
\end{equation}
involving a generalized diffusion coefficient $D(\omega)$ renormalized by quantum interference according to:
\begin{equation}
\label{eq:SCTL_D}
\frac{1}{D(\omega)}=\frac{1}{D_B}+\frac{1}{\pi\rho\hbar D_B}
\int \frac{\dd \boldsymbol{Q}}{(2\pi)^3}
\frac{1}{-i\omega+D(\omega)\boldsymbol{Q}^2},
\end{equation}
where $D_B$ is the so-called Boltzmann or classical diffusion coefficient and $\rho$ the density of states. An exact calculation of the frequency-dependent diffusion coefficient $D(\omega)$ and the corresponding density $n(\br,t)$ was previously performed in \cite{Lobkis2005} in one and two dimensions. Here, we extend this analysis to the 3D case. 
To this end, we first note that the integral on the right-hand side of Eq. (\ref{eq:SCTL_D}) is divergent and must be regularized by introducing a UV cutoff $Q_\text{UV}$, typically of the order of the inverse mean free path. This can be conveniently achieved by isolating the divergence in the momentum integral as follows \cite{Cherroret2025}:
\begin{align}
\int \frac{\dd Q}{2\pi^2 D(\omega)}\frac{Q^2\!-\!i\omega/D(\omega)\!+\!i\omega/D(\omega)}{-i\omega/D(\omega)+Q^2}=
\frac{1}{2\pi^2}\left[
\int_0^{Q_\text{UV}}\!\!\!\frac{\dd Q}{D(\omega)}\!+\!\frac{i\omega}{D(\omega)}\int_0^\infty\!\frac{\dd Q}{-i\omega/D(\omega)\!+\!Q^2}\right].
\end{align}
Carrying out the two integrals yields a third-order algebraic equation for
 $D(\omega)$, whose solution is 
\begin{equation}
\label{Eq:Dsol}
D(\omega)=\left[\beta
\frac{-\alpha_0(1-E/\Ec)+[\sqrt{\alpha_0^3(1-E/\Ec)^3-i\omega}+\sqrt{-i\omega}]^{2/3}}{[\sqrt{\alpha_0^3(1-E/\Ec)^3-i\omega}+\sqrt{-i\omega}]^{1/3}}\right]^2,
\end{equation}
where $\beta=(8\pi^2\rho\hbar)^{-1/3}$ and $\alpha_0(1-E/\Ec)=D_B[Q_\text{UV}/(2\pi^3\rho D_B\hbar)-1]/(3\beta^2)$ (restoring $\hbar$).
In this formulation, the three physical quantities $\rho$, $D_B$ and $Q_\text{UV}$ of the model are recast in terms of the three composite parameters $\alpha_0$, $\beta$ and $\Ec$. The latter, in particular, represents the mobility edge. 
While these parameters can be approximately estimated using perturbation theory \cite{Akkermans07, Kuhn2007, Richard2019}, they are difficult to determine accurately in the strong-disorder regime where the Anderson transition occurs. Here we will therefore simply take them as constant free parameters of the theory, exploiting  that they vary smoothly with  energy $E$ in the vicinity of the mobility edge we are focusing on.
%Yedjour2010,
Together, Eqs. (\ref{eq:energy_dist}), (\ref{eq:density_full}), (\ref{eq:Pqomega}) and (\ref{Eq:Dsol})  provide a complete description of the wave-packet density at any position and time across the critical point. It is nevertheless instructive, at this stage, to recall the predictions of the theory in the limit of asymptotically long times, corresponding to $\omega\to0$ \cite{Shapiro1982}:
\begin{equation}
\label{eq:Dasymptotics}
    D(\omega)\underset{\omega\to0}{\simeq}
    \begin{cases}
        3\beta^2\alpha_0\left(\dfrac{E}{\Ec}-1\right)\equiv D_0&E>\Ec    \vspace{0.15cm}\\
   \vspace{0.1cm}
   \beta^2(-4i\omega)^{1/3}& E=\Ec\\
        -i\omega\dfrac{4\beta^2}{9\alpha_0^2(1-E/\Ec)^2}\equiv -i\omega\xi^2&E<\Ec
    \end{cases}
\end{equation}
In terms of wave-packet spreading, these  frequency scaling laws imply fundamentally different dynamics around the mobility edge, which can be traced in the time evolution of the mean squared width of the wave packet, $\langle\br^2(t)\rangle\equiv
    \int \dd \br\, \br^2 n(\br,t)$. 
    Assuming a sharp filter function, so that the wave packet has a well defined energy $\Ef$, $\langle\br^2(t)\rangle$ readily follows from Eqs. (\ref{eq:density_narrow}) and (\ref{eq:Pqomega}):
\begin{align}
    \langle\br^2(t\to\infty)\rangle\simeq
    -\int\frac{\dd \omega}{2\pi}
    \nabla_\bq\cdot\nabla_\bq\frac{e^{-i\omega t}}{-i\omega+D(\omega\to0)\bq^2}\Big|_{\bq=0}=
    \int\frac{\dd \omega}{2\pi}
    \frac{6D(\omega\to0)e^{-i\omega t}}{(-i\omega+0^+)^2}.
\end{align}
Inserting the asymptotic laws (\ref{eq:Dasymptotics}), we infer:
\begin{equation}
\label{eq:longtime_scaling}
   \langle\br^2(t\to\infty)\rangle\propto
   \begin{cases}
       t & E>\Ec\\
       t^{2/3} &E=\Ec\\
       \text{const.} & E<\Ec,
   \end{cases}
\end{equation}
which are the usual diffusive, anomalous and localized behavior expected above, at and below the mobility edge \cite{Abrahams1979, ohtsuki1997anomalous}.

\section{Benchmarking the theory against numerical simulations}
\label{Sec:numerics}
\vspace{0.3cm}

Before analyzing the experimental results of Ref.~\cite{Josse2026} within the theoretical framework introduced in the previous section, we first assess its validity through comparison with \emph{ab initio} numerical simulations. These simulations are performed in an idealized setting assuming sharp energy filtering and a spatially narrow initial state, yet under conditions close to those of the experiment.
To this end, we numerically prepare, in a blue-detuned speckle potential $V(\mathbf{r})$, an initial filtered state $|\psi_0\rangle$ of the form given in Eq.~(\ref{eq:expansion_psi0}), and then compute its time evolution $\psi(\br,t)$ in the random potential. The density $|\psi(\br,t)|^2$ is monitored as a function of time and subsequently averaged over disorder realizations, yielding the average density $n(\br,t)\equiv\overline{|\psi(\br,t)|^2}$.

\subsection{Speckle potential}
\vspace{0.3cm}
\label{Sec:Speckle potential}
The speckle potential $V(\br)$ is generated numerically on a regular rectangular grid of volume $(\SI{100}{\micro \meter})^3$, using a Fourier filtering method similar to the one described in Ref.~\cite{Volchkov2018}. The disorder corresponds to a blue-detuned laser speckle, as in Ref.~\cite{Josse2026}, implying that the potential is strictly positive. Its statistical distribution follows the exponential law $P(V) = \exp( -V/\VR)/\VR \cdot \Theta(V)$~\cite{Clement2006}. Here, $\VR$ denotes the root-mean-square (rms) value of the disorder and is referred to as the disorder amplitude. For laser speckle potentials, $\VR$ also coincides with the mean value of the disorder, i.e., $\overline{V} = \VR$. In the following we present numerical results for $\VR/h=\SI{416}{\hertz}$, as in Ref.~\cite{Josse2026}.

\begin{figure}[t]
\centering
\includegraphics[width = 0.292\textwidth, trim=1.5cm .4cm 2.7cm 0.4cm, clip]{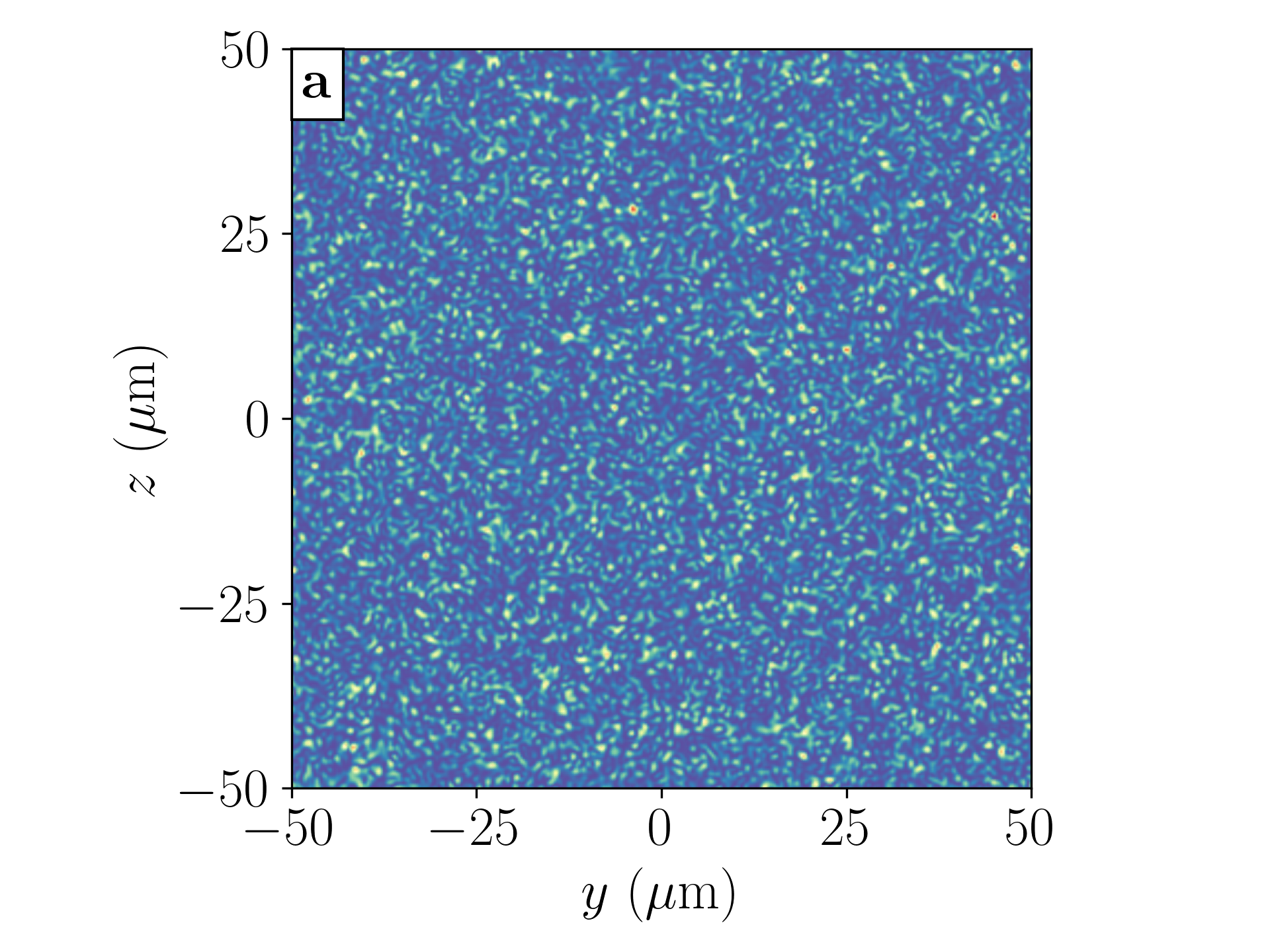}
~
\includegraphics[width = 0.31\textwidth, trim=2.85cm .4cm .6cm 0.4cm, clip]{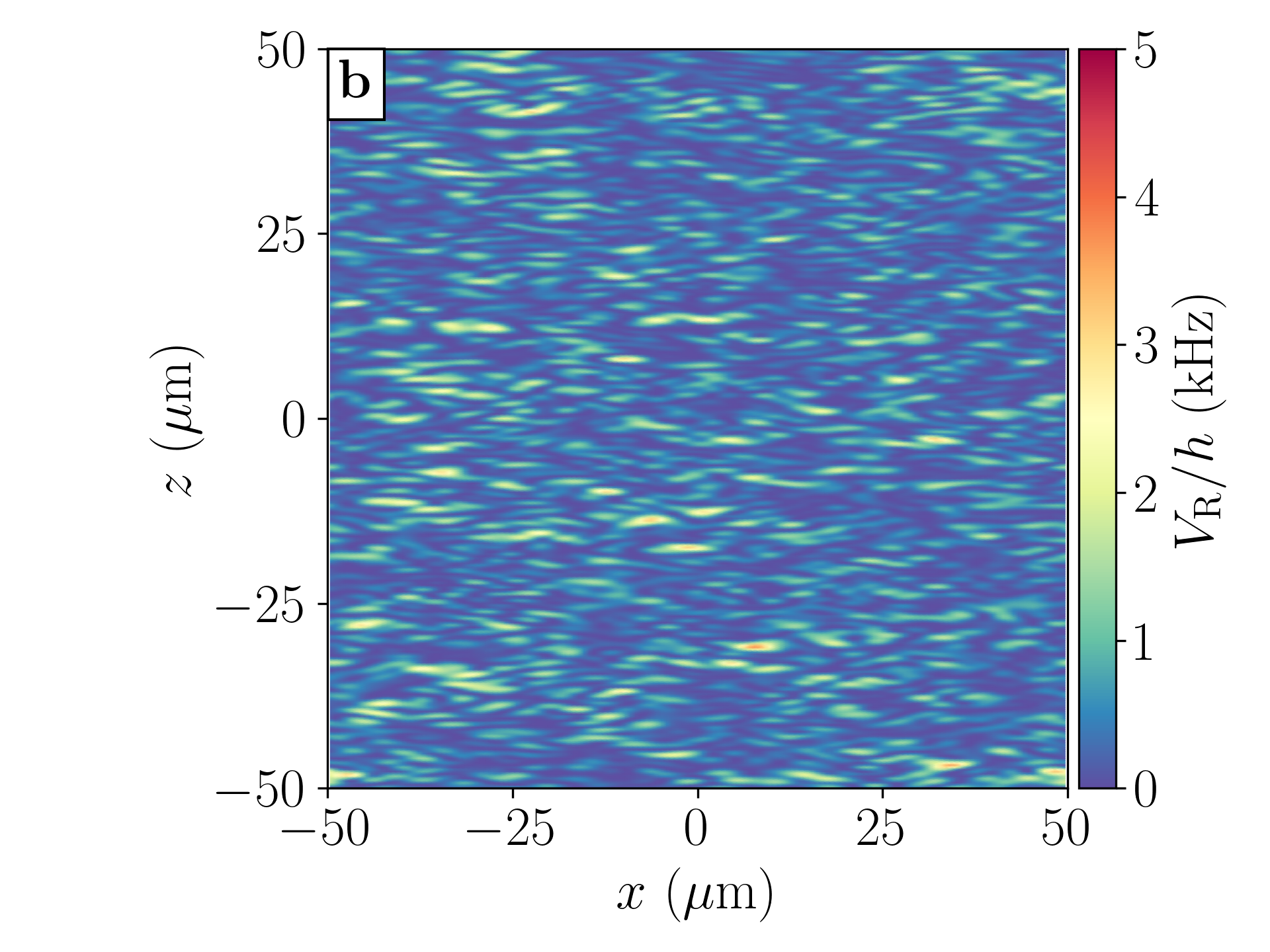}
~
\includegraphics[width = 0.35\textwidth, trim=.4cm 0.cm .4cm 0.3cm, clip]{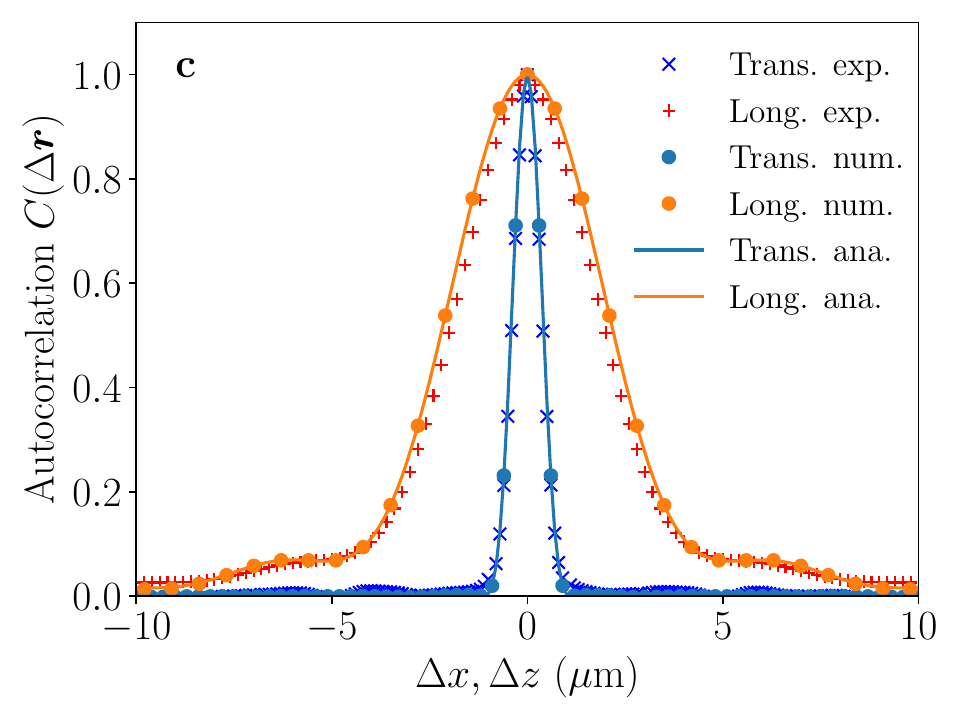}
\vspace*{-.5cm}
\caption{\justifying (a) and (b) Two-dimensional cuts of a realization of a three-dimensional speckle potential, and (c) its autocorrelation function along the longitudinal ($x$) and a transverse ($z$) directions. The speckle  parameters are: $\VR/h = 416$~Hz,  $\sigma_x = \SI{1.45}{\micro \meter}$,  and $\sigma_\perp = \SI{0.30}{\micro \meter}$.  Crosses: experimental data; points: numerical data; solid lines: analytical autocorrelation function, Eq.~\eqref{eq:autocorr}. See Appendix~\ref{App:speckle} for further details.}
\label{Fig:speckle}
\vspace*{-.3cm}
\end{figure}

The speckle potential is generated to accurately reproduce the experimental configuration, where the speckle grains are elongated along the $x$ axis (see Fig.~\ref{Fig:speckle}). More precisely, the parameters are chosen such that the autocorrelation function $C(\Delta \br) = \overline{\delta V(\br)\delta V(\br + \Delta \br)} /\VR^2$ (where $\delta V\equiv V-\VR $) matches the experimental one both along the longitudinal $x$ axis and in the transverse $y$–$z$ plane.
 As shown in Fig.~\ref{Fig:speckle}c, the agreement between the numerical computation, the theoretical expression given by Eq.~\eqref{eq:autocorr} in the Appendix~\ref{App:speckle}, and the experimental measurements~\cite{Josse2026, Volchkov2018} of the autocorrelation, is excellent. Consistently with Ref.~\cite{Pasek2017}, the correlation lengths are defined as the half width at half maximum (HWHM) divided by a factor 1.39156. It yields $\sigma_x = \SI{1.45}{\micro \meter}$ and $\sigma_\perp = \SI{0.30}{\micro \meter}$ on the transverse plane. These  correlation lengths set the correlation energy $E_\sigma = \hbar^2 / m \sigma^2 = h \times \SI{441}{\hertz}$, where $\sigma = \sigma_x^{1/3} \sigma_\perp^{2/3} = \SI{0.50}{\micro \meter}$ is the geometric mean over the different directions.  

In the simulations, the discretization steps are typically $\Delta x = \SI{0.7}{\micro \meter}$ and $\Delta y = \Delta z =\SI{0.3}{\micro \meter}$, yielding $\sigma_x/\Delta x \approx 2$, $\sigma_\perp/\Delta z \approx 1$. These ratios are sufficient to properly resolve both the speckle grains [see Fig.~\ref{Fig:speckle}(c)] and the de Broglie wavelength at the considered energies. Details of the numerical generation of the speckle potential and the theoretical form of the autocorrelation function are provided in Appendix~\ref{App:speckle}.

\subsection{Numerical propagation of an energy-resolved wave-packet}
\vspace{0.3cm}

\sloppy To numerically prepare the initial state $|\psi_0\rangle$ defined by Eq. (\ref{eq:expansion_psi0}), we start from a Gaussian wave packet $\phi (\br) \propto \exp\big( - \br^2/4\Delta r^2 \big)$, and consider a narrow Gaussian filter $\smash{f(E-\Ef) \propto \exp\big[-(E-\Ef)^2/ 4\Delta E^2\big]}$, with  $\Delta r = \SI{3.5}{\micro \meter}$ and $\Delta E/h = 5$~Hz. %= 3.898$~Hz.
The target energy $\Ef$ is varied around the expected mobility edge, typically from $\Ef/h=166$ to $346$~Hz. To apply the  filtering operation $\smash{f(\hat{H}-\Ef)}$ involved in Eq. (\ref{eq:expansion_psi0}), we do not compute the eigenstates of the random Hamiltonian, which  would be numerically prohibitive for the 3D system considered here. Instead, we use an expansion of the operator $f(\hat{H}-\Ef)$ on a basis of Chebyshev polynomials. This strategy was previously employed in Refs. \cite{Ghosh2014, Ghosh2015, Ghosh2017} in the context of Anderson localization in momentum space, and is briefly described in Appendices~\ref{App:chebyshev} and \ref{App:filter}.

Once the energy-resolved state $\Ket{\psi_0}$ is computed following Eq. (\ref{eq:expansion_psi0}),  we numerically propagate it in time by applying the time evolution operator $\Ket{\psi(t)} = \exp \big(-i\hat{H} t \big) \Ket{\psi_0}$. The time propagation is performed with high accuracy and efficiency using a Chebyshev expansion of the evolution operator, which is applied iteratively to obtain the state at successive time steps $\Delta t$.  This method, briefly reviewed in Appendices~\ref{App:chebyshev} and~\ref{App:time_evolution}, is more stable and efficient than Crank–Nicolson schemes, as it allows for relatively large time steps \cite{Fehske2009, Roche1997}. Within this approach, we obtain the spatial density at time $t$, which we average over realizations of the disorder to get the average density $n(\br,t)$.

\begin{figure}[t]
 \centering
 \includegraphics[width = 0.48\textwidth, trim=.35cm .3cm .35cm 0.3cm, clip]{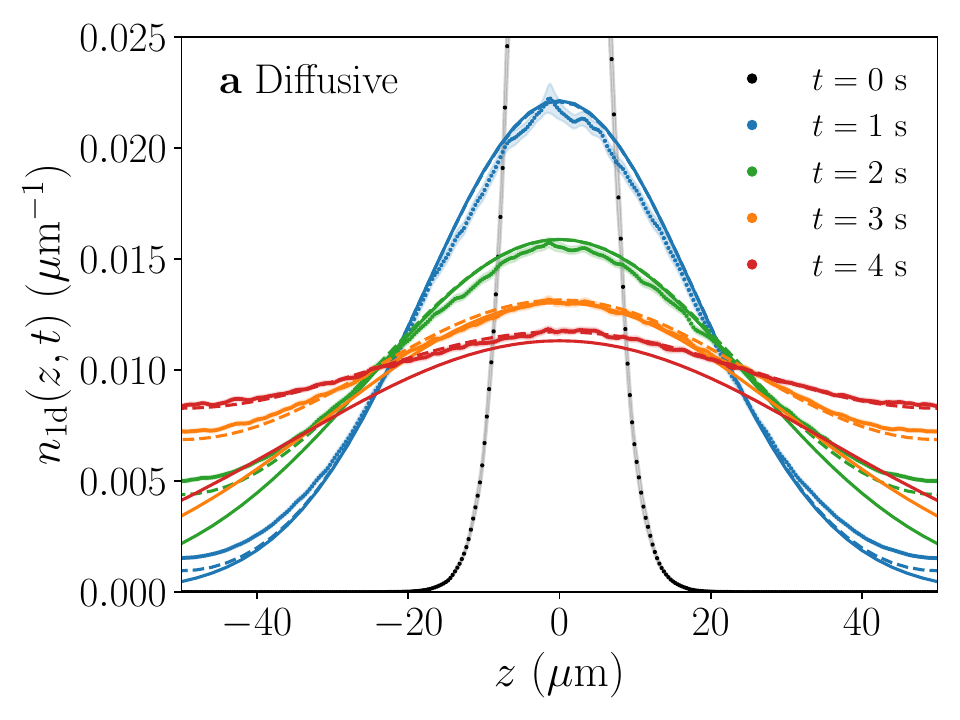}
 ~
 \includegraphics[width = 0.48\textwidth, trim=.35cm 0.3cm .35cm 0.3cm, clip]{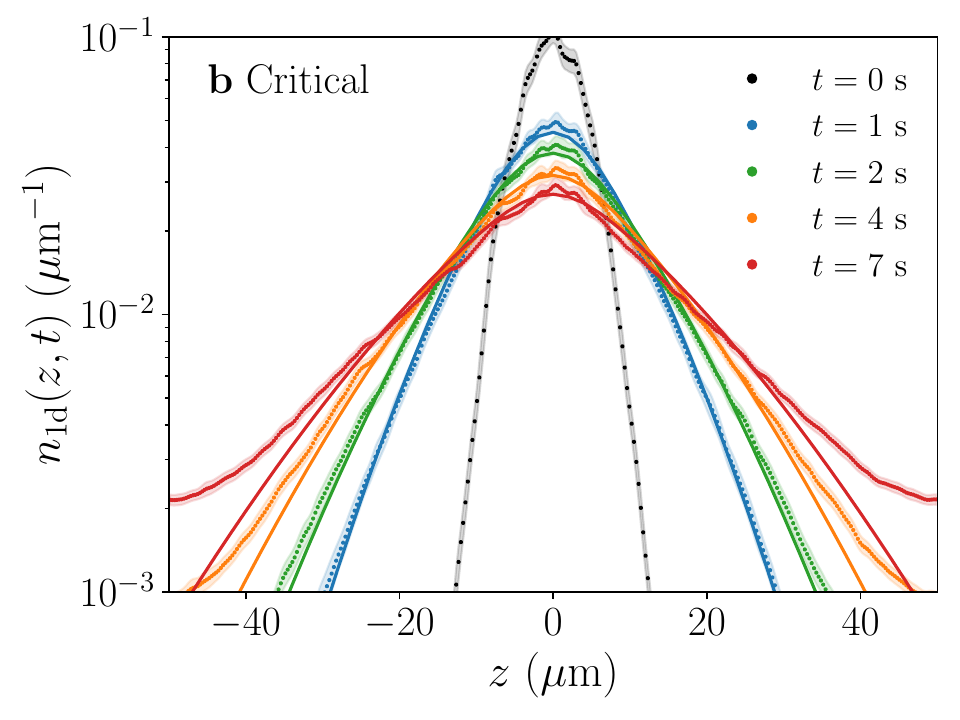}
 
 \includegraphics[width = 0.48\textwidth, trim=.35cm 0.3cm .35cm 0.3cm, clip]{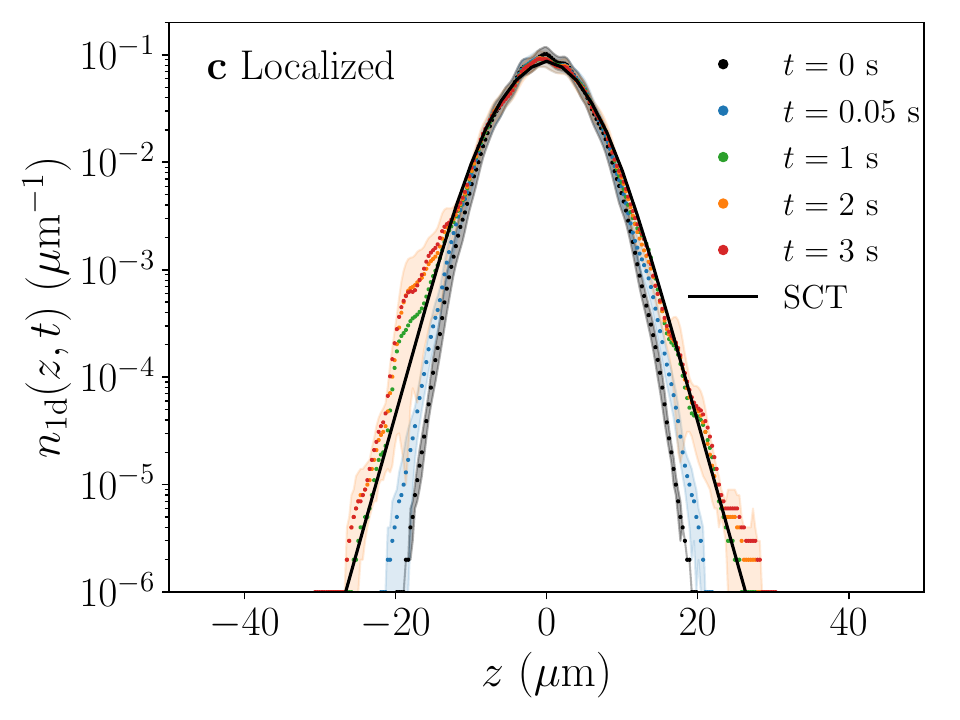}
 ~
 \includegraphics[width = 0.48\textwidth, trim=.35cm 0.3cm .35cm 0.3cm, clip]{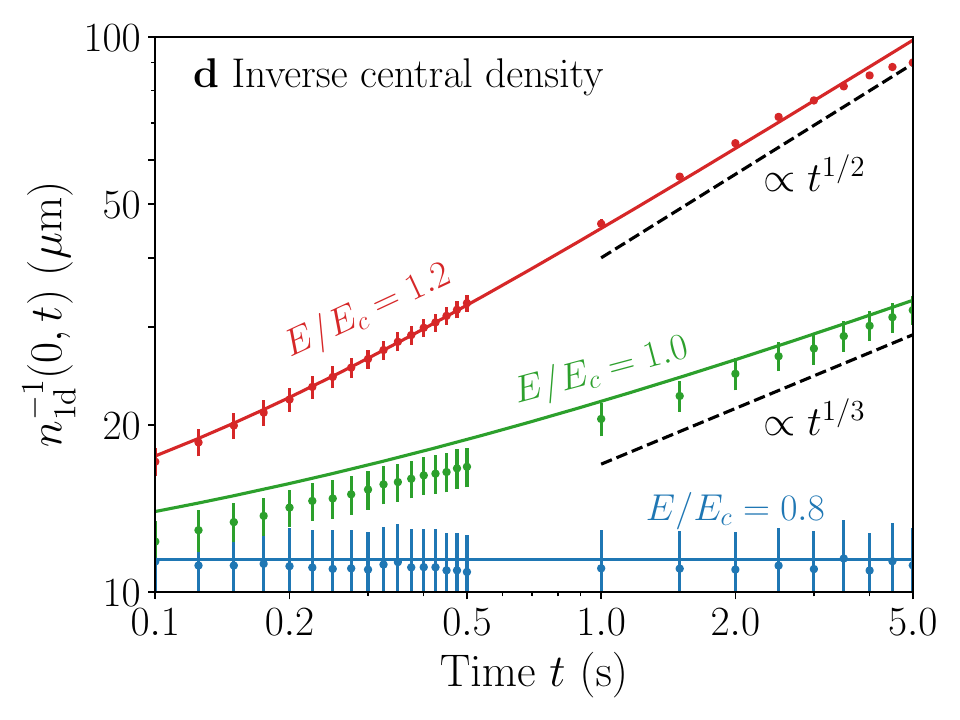}
 \caption{\justifying
 (a)-(c) Linear density profiles $n_\text{1d}(z,t)$ across the 3D Anderson transition in a speckle potential, obtained from numerical simulations of wave-packet propagation (discrete points) for a disorder characterized by the properties shown in Fig. \ref{Fig:speckle}. Three target energies are considered: (a)$\Ef/h=326$~Hz, (b) $\Ef/h=266$~Hz, and (c) $\Ef/h=206$~Hz, corresponding respectively to the diffusive, near critical, and localized regimes. The solid lines show the long-time density profiles predicted by the SCT. In the diffusive regime (a), the colored dashed curves show the theoretical profiles including boundary corrections, Eq. (\ref{eq:profile_finiteL}). 
 (d) Inverse of the central density as a function of time for $\Ef/h=326$~Hz ($E/\Ec\simeq 1.2$), $\Ef/h=266$~Hz ($E/\Ec\simeq 1$), and $\Ef/h=206$~Hz ($E/\Ec\simeq 0.8$), revealing the three characteristic long-time scaling laws (\ref{eq:longtime_scaling}) across the mobility edge. Points are the results of numerical simulations and solid lines show the predictions of the SCT. Dashed lines are guides indicating the algebraic laws expected at long time. Here numerical results are averaged over 160 disorder realizations. The shaded regions and error bars around the data points correspond to $\pm 2$ standard deviations on the average. We fit $(\alpha_0,\beta)\simeq (8.74 \pm 0.24\,\text{s}^{-1/3}, 4.93 \pm 0.06\, \mu \text{m}.\text{s}^{-1/3})$ for the SCT parameters. Uncertainties on the parameters are derived from the fitting procedure and given as $\pm 2$ standard errors.
 }
\label{Fig:profiles}
\end{figure}

To characterize the dynamics, we follow the methodology of Ref.~\cite{Josse2026} and we focus on the disorder-averaged linear density profile along $z$, defined as $n_{\mathrm{1d}} (z,t) = \iint \dd x \dd y \: n(\br,t)$. Figure~\ref{Fig:profiles}a–c shows such linear density profiles at different evolution times obtained from the numerical simulations (colored points) for three target energies $\Ef$, corresponding respectively to the diffusive ($\Ef>\Ec$),  critical ($\Ef=\Ec$) and localized regime ($\Ef<\Ec$). For the chosen numerical parameters, we estimate the mobility edge to be $\Ec \simeq 263~\mathrm{Hz}$.
This value slightly exceeds the numerical estimate $\Ec \simeq 240~\mathrm{Hz}$ reported in Ref.~\cite{Pasek2017}, which is based on extensive transfer-matrix simulations. We attribute this discrepancy to the finite size of the numerical grid, which restricts the accessible expansion times, and to the relatively coarse spatial discretization required to keep the computational cost of long-time expansion in a 3D disordered potential tractable. 
Despite this, the simulation results clearly reveal the characteristic Gaussian spreading and the time-independent exponential profile expected on the diffusive (Fig. \ref{Fig:profiles}a) and localized (Fig. \ref{Fig:profiles}c) sides of the Anderson transition, respectively. Near the mobility edge (Fig. \ref{Fig:profiles}b), the profile exhibits an intermediate shape together with a slow expansion.

\subsection{Comparison with the 3D self-consistent theory}
\label{Sec:comparison_num_sct}
\vspace{0.3cm}

To compare the numerical results of Fig. \ref{Fig:profiles} with the SCT, we start from the narrow-filter approximation (\ref{eq:density_narrow}) applied to the linear density, 
\begin{equation}
 \label{eq:linear_density}
    n_\text{1d}(z,t)\simeq 
    %\int dE |f(E)|^2 A(E,\bk=0)
    \int_{-\infty}^\infty \dd z' P_{E}(z,z',t) n_\text{1d}(z',t=0)\;,
\end{equation}
where $n_\text{1d}(z',t=0)\equiv \int dx' dy' |\phi(x',y',z')|^2$. The 1D propagator is deduced from Eq.~\eqref{eq:Pqomega}:
\begin{align}
\label{eq:P}
P_{E}(z,z',t)&=\int_{-\infty}^\infty \frac{\dd \omega}{2\pi}\int_{-\infty}^\infty \frac{\dd q_z}{2\pi}\frac{e^{iq_z(z-z')-i\omega t}}{-i\omega+D(\omega)q_z^2}\\ \nonumber
&=\int_{-\infty}^\infty \frac{\dd \omega}{2\pi}\frac{e^{-i\omega t}}{\sqrt{-4 i\omega D(\omega)}}\exp\Big[\!-\!|z-z'|\sqrt{\frac{-i\omega}{D(\omega)}}\Big],
\end{align}
with the diffusion coefficient $D(\omega)$ determined from Eq.~\eqref{Eq:Dsol}.
Note that the propagator is normalized, $\int \dd z P_E(z,z',t)=1$, and so is the linear density $n_\text{1d}(z,t)$. 
Because the profiles represented in Fig. \ref{Fig:profiles} typically correspond to times well beyond to the mean free time $\tau$\footnote{
From \cite{Richard2019}, we estimate a scattering mean free time $\tau\simeq 1\,$ms for a speckle disorder with the properties of Fig. \ref{Fig:speckle}.
}, Eq.~\eqref{eq:P} can be further simplified by using the small-frequency asymptotic relations \eqref{eq:Dasymptotics} for $D(\omega)$. In the diffusive regime $E>\Ec$, they lead to the well-known Gaussian distribution
\begin{align}
\label{eq:asymp_diff}
    P_{E>\Ec}(z,z',t)=\frac{1}{\sqrt{4\pi D t}}\exp\left(-\frac{|z-z'|^2}{4Dt}\right)
\end{align}
where $D=3\beta^2\alpha_0(E/\Ec-1)$, while in the localized regime $E<\Ec$ one has
\begin{align}
\label{eq:asymp_loc}
    P_{E<\Ec}(z,z',t)=\frac{1}{2\xi}\exp\left(-\frac{|z-z'|}{\xi}\right)
\end{align}
where $\xi=2\beta/[3\alpha_0(1-E/\Ec)]$. At the mobility edge $E=\Ec$, finally, the linear density is an Airy function \cite{Lemarie2009, Akridas2019}:
\begin{align}
\label{eq:asymp_ME}
    P_{E=\Ec}(z,z',t)=\frac{3\sqrt{\varrho}}{2(3t)^{1/3}}
    \text{Airy}\left[\frac{\sqrt{\varrho}|z-z'|}{(3t)^{1/3}}\right]
\end{align}
with $\varrho=4^{-1/3}\beta^{-2}$.

Figure~\ref{Fig:profiles}a-c shows as solid curves the SCT prediction \eqref{eq:linear_density} for the linear density, using the asymptotic expressions \eqref{eq:asymp_diff}, \eqref{eq:asymp_loc} and \eqref{eq:asymp_ME}. For the comparison, we use the average numerical profiles immediately after loading into the disorder as the initial density distribution%\footnote{\textcolor{blue}{Needs modification : In the derivation of Eq.~(\ref{eq:linear_density}), it is assumed that the BEC profile $|\phi(\mathbf{r})|^2$ in state $|1\rangle$ remains unchanged by the rf transfer. Although we have verified numerically that this assumption holds to a good approximation, a slightly more accurate comparison is obtained by using in Eq.~(\ref{eq:linear_density}) the post-transfer profile, of size $\Delta z\simeq4.6\,\mu$m, which is slightly larger than $\Delta r=3.5\,\mu$m.}}
, and take $\alpha_0$ and $\beta$ as fit parameters.
These are calibrated from a global comparison between the SCT and the numerical profiles. To ensure consistency when comparing the SCT to both the numerical and experimental profiles (see Sec.~\ref{Sec:comparison_exp_sct}), we use the central density $n_\text{1d}(z=0,t)$ as the fitting observable. This is motivated experimentally since the central density is less noisy than, for example, the tails of the profiles (which are affected by finite signal-to-noise ratio), while still fully capturing the dynamics of the cloud expansion. The parameters $\alpha_0$ and $\beta$ are determined by minimizing the least-squares cost function
\begin{equation}
\chi^2(\alpha_0, \beta) = \sum_{\Ef} \sum_t \left[ \frac{n_\text{1d}(0,t)|_{\Ef} - n_{\text{SCT}}(0,t)|_{\Ef,\alpha_0,\beta}}{\sigma(0,t)|_{\Ef}}\right]^2,
\label{Eq:optimization}
\end{equation}
where $\smash{n_\text{1d}(0,t)\big|_{\Ef}}$ denotes here the numerical central density at energy $\Ef$, $\sigma(0,t)|_{\Ef}$ its associated uncertainty and $\smash{n_{\text{SCT}}(0,t)\big|_{\Ef,\alpha_0,\beta}}$ the corresponding SCT prediction for a given parameter pair $(\alpha_0,\beta)$.
The double sum runs over target energies spanning the mobility edge and over evolution times up to five seconds. The minimization yields the values $(\alpha_0,\beta) = (8.74 \pm 0.24\,\text{s}^{-1/3}, 4.93 \pm 0.06\, \mu \text{m}.\text{s}^{-1/3})$, where the quoted uncertainties correspond to the fitting errors.

Overall, Fig.~\ref{Fig:profiles}a–c shows very good agreement between the SCT and the \emph{ab initio} simulations over the range of times and energies explored (typically $\Ef = E_c \pm 0.2E_c$, for evolution times up to $7\,$s). In particular, the tails of the density profiles are very well reproduced by the theory, both in the localized regime and at the critical point—an encouraging result given the approximate nature of the SCT. Note that despite the relatively narrow initial density distribution, the profiles remain close to the initial shape in the localized regime for the chosen disorder amplitude. This behavior is also observed experimentally (see Sec.~\ref{Sec:experiments}).

In the diffusive regime, by contrast, we observe significant deviations between theory and simulations in the tails. These discrepancies originate from the finite size $L$ of the numerical domain, which starts to affect the profiles once the wave packet reaches the domain boundaries.
This interpretation is supported by theory: replacing the unbounded expression (\ref{eq:asymp_diff}) with the diffusive solution in a finite box of size $L$ (with periodic boundary conditions, as in the simulations), we find \cite{Akkermans07}
\begin{align}
\label{eq:profile_finiteL}
n_\text{1d}(z,t)=
\frac{1}{L}
\left[1+2\sum_{n=1}^\infty
\cos\Big(\frac{2\pi n z}{L}\Big)
\exp\Big[-\Big(D t+\frac{\Delta z^2}{2}\Big)\Big(\frac{2\pi n}{L}\Big)^2\Big]
\right],
\end{align}
which yields an excellent agreement with the simulations, including in the tails (see Fig. \ref{Fig:profiles}a).

Figure \ref{Fig:profiles}d finally shows the inverse central density of the wave packet, $n^{-1}_\text{1d}(z=0,t)$, as a function of time for three target energies in the diffusive, localized, and critical regimes. The plot reveals the three characteristic long-time scaling laws (\ref{eq:longtime_scaling}) for the packet width,
$\sqrt{\langle\mathbf{r}^2(t)\rangle}\propto \sqrt{\langle z^2(t)\rangle}\propto n^{-1}_\text{1d}(0,t)$, with again very good agreement between the SCT and the \emph{ab initio} simulations. This analysis confirms that the theoretical framework reliably captures the dynamics across the mobility edge, even in a regime of relatively strong disorder, large spatial scales, and long times, where direct numerical simulations become highly demanding.

%\vspace{0.3cm}
%\begin{figure}[H]
% \centering
% \includegraphics[scale=0.35]{Inverse_Maximum.pdf}
% \caption{
% (a) Inverse linear density as a function of energy at increasing times. (b) Inverse linear density as a function of time for three energies below, at and above the mobility edge. Points are numerical  simulation results, and solid curves are f its to the self-consistent theory.  \textcolor{red}{to complete and modify}
%  }
%\label{Fig.inverse_density}
%\end{figure}

\section{Theory versus experiments}
\label{Sec:experiments}
\vspace{0.3cm}

We now present comparisons between the SCT and the experimental observation of the density-profile dynamics obtained with the setup of Ref.~\cite{Josse2026}. We begin with a brief description of the experimental conditions (Sec. \ref{Sec:exp_description}) and of the corresponding initial density profiles, which serve as the input to the SCT. We then characterize the atomic energy distribution realized in the experiment, accounting for the residual thermal background cloud (Sec.~\ref{Sec:exp_edistribution}). Finally, we provide a quantitative comparison between the measured density profiles and the theoretical predictions over a range of target energies spanning the mobility edge.

\subsection{Energy-resolved state preparation}
\label{Sec:exp_description}
\vspace{0.3cm}

The crucial feature of the experiments reported in Ref.~\cite{Josse2026} is the energy-resolved state preparation, as sketched in Fig.~\ref{Fig:scheme}. The  sequence starts with the preparation of a $^{87}\mathrm{Rb}$ BEC in state $|1\rangle=|F=2,m_F=1 \rangle$, containing around $2\times 10^5$ atoms in a nearly isotropic optical trap with frequencies $(\omega_x, \omega_y, \omega_z) \approx 2\pi \times (35, 25, 11)$~Hz. The temperature is $T \simeq 7~\mathrm{nK}$ and the condensate fraction is around $\fc\simeq75\%$. The BEC is  characterized by the chemical potential $\mu_{\rm in}/h \simeq 350~\mathrm{Hz}$, yielding the Thomas–Fermi radii $R_{\rm TF} \simeq {(8, 12, 26)}$~\textmu m.

The state-dependent disorder is created using a bichromatic optical speckle, following the principle described in Ref.~\cite{Lecoutre2022}. A small fraction of atoms ($\sim 5\%$) is then transferred to the state $|2\rangle= |F=1,m_F=-1 \rangle$ by applying a radio-frequency (rf) pulse. The rf frequency sets the target energy to $\Ef=\hbar \delta_{\rm rf}$, where $ \delta_{\rm rf}$ is the detuning from the bare resonance in absence of disorder (see Fig.~\ref{Fig:scheme}). The rf pulse duration, $t_{\rm rf}\simeq 40$~ms, defines the energy width $\Delta E\sim h/t_{\rm rf}$ of the energy-resolved wave packet  (see below). Immediately after the transfer, at $t=0$, the optical trap is switched off and the remaining atoms in state $|1\rangle$ are removed. The subsequent evolution of the wave packet in the disordered potential is then monitored up to 5 seconds by \emph{in-situ} fluorescence imaging along the $x$ axis. During the whole experimental sequence, the atoms are suspended against gravity using a magnetic levitation scheme. Such levitation yields a shallow residual harmonic confinement\footnote{Although the precise influence of the weak confinement remains to be precisely investigated, the transport dynamics remains fully 3D since the transverse size of the cloud is much larger than the expected scattering mean path $\ell\simeq\SI{1}{\micro \meter}$, see Ref.~\cite{Josse2026}.} in the horizontal $x$-$y$ plane, such that the expansion occurs essentially along the vertical $z$ axis. Following the analysis done in Ref.~\cite{Josse2026}, we then integrate further the recorded column density $n_{\rm col}(y,z,t)=\int \dd x\; n_{\rm exp}(\br, t)$, and we study the one-dimensional profile $n_{\rm 1d}(z,t)=\int \dd y\, n_{\rm col}(y,z,t)$ along $z$.

\begin{figure}[t]
\centering
\includegraphics[width=\textwidth,trim=0.25cm .25cm 0.25cm 0.25cm,clip]{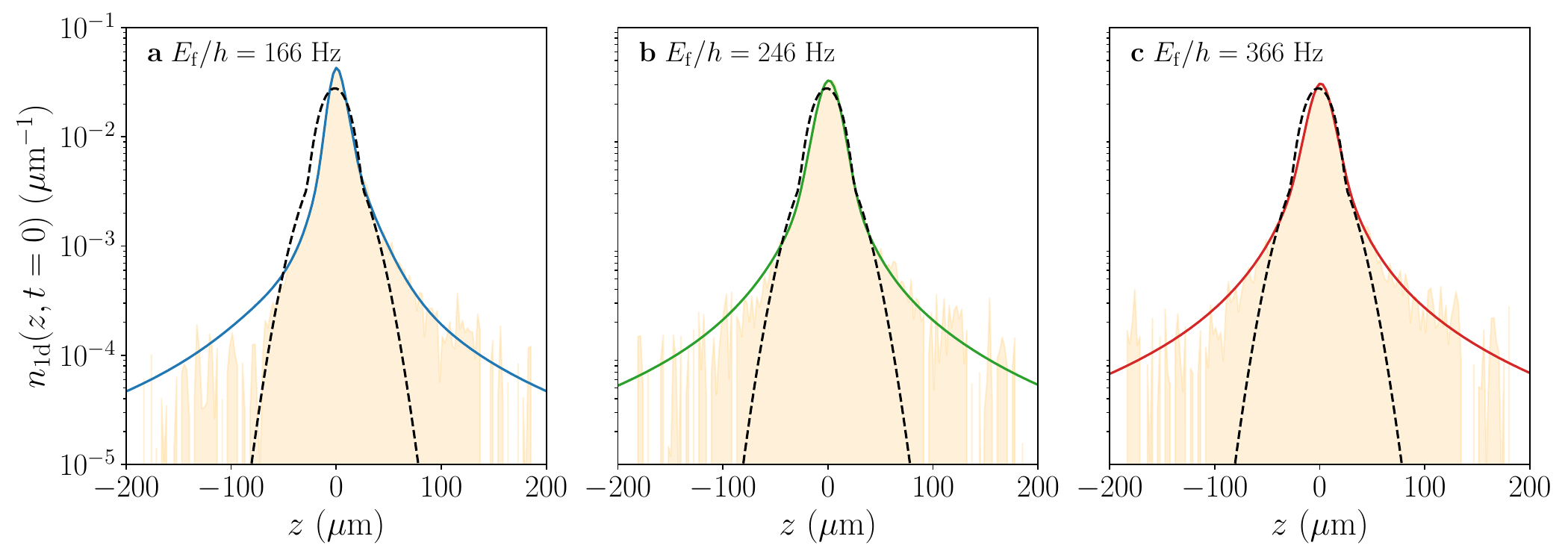}
\vspace*{-.7cm}
\caption{\justifying Initial one-dimensional density profiles of atoms in state $|2\rangle$ immediately after transfer along the $z$ axis, $n_{\rm 1d}(z,t=0)$.
The shaded orange areas show the experimental profiles on a semi-log scale for three target energies: (a) $\Ef/h=166$, (b) $246$, and (c) $366$~Hz. Dashed curves indicate, for comparison, the time-of-flight estimate of the BEC density profile in state $|1\rangle$ before the rf transfer, including the thermal component. Both profiles are normalized to unit integral for proper comparison. Solid curves are fits of $n_{\rm 1d}(z,t=0)$ to pseudo-Voigt profiles, which are used as input for the SCT.}
\label{fig:Initialprofile_NJP}
\end{figure}

For each energy $\Ef$, we record the initial profile $n_{\rm 1d}(z,t=0)$ immediately after transfer. Examples are shown in Fig.~\ref{fig:Initialprofile_NJP} for $\Ef/h=166$, $246$ and $366$~Hz (solid curves with orange shading). The dashed lines indicate the estimated initial density profile of the BEC in state $|1\rangle$ before transfer, including the thermal component. The experimental profiles $n_{\rm 1d}(z,t=0)$ exhibit weak tails, which we attribute to a combination of short-time interaction effects and possible dynamics in state $|2\rangle$ during the rf-coupling stage. They also display a slight asymmetry, likely caused by small fluctuations and inhomogeneities in the levitation magnetic field.

\subsection{Energy distribution}
\label{Sec:exp_edistribution}
\vspace{0.3cm}
The key objective of the rf transfer scheme is to prepare an atomic cloud with a narrow energy distribution in state $|2\rangle$. In the weak-coupling regime in which the experiment operates~\cite{Josse2026}, the energy spread of the transferred atoms is Fourier-limited, $\Delta E \sim h/t_{\rm rf}$. In this case, and assuming that the atoms in state $|1\rangle$ form a pure BEC, the transferred state $|\psi_0^{\rm exp}\rangle$ is also pure and has an energy distribution $\mathcal{D}_{\rm BEC}(E;\Ef)=F(E-\Ef)\, A(E,\bk=0)$ given by Eq.~\eqref{eq:energy_dist}. The filter function $F(E)$ is determined by the Fourier transform of the temporal envelope of the rf pulse, which has the shape of a Kaiser function~\cite{Josse2026}. This leads to a narrow rms energy width $\Delta E/h \simeq 14~\mathrm{Hz}$.

In the experiment, however, the initial atomic cloud in state $|1\rangle$ is not a fully pure BEC but it contains a finite thermal fraction, $1-\fc\simeq 25\%$. This thermal component gives rise to an additional contribution $\mathcal{D}_{\rm th}$ in the energy distribution, which generalizes to:
\begin{equation}
\label{eq:Dtotal}
\mathcal{D}(E;\Ef)= \fc \; \mathcal{D}_{\rm BEC}(E;\Ef)+ (1-\fc) \; \mathcal{D}_{\rm th}(E;\Ef),
\end{equation}
where (see Appendix~\ref{App:energy_distr})
\begin{align}
\label{eq:Dthermal}
\mathcal{D}_{\rm th}(E;E_{\rm f}) & \simeq \frac{2(k_\mathrm{B} T)^{-3/2}}{\sqrt{\pi}}\,
\Theta(E-\Ef)\,\sqrt{E-\Ef}\, \exp\left( -\frac{E-\Ef}{k_\mathrm{B}T} \right) \; .
\end{align}

\begin{figure}[t]
\centering
\includegraphics[width=0.65\textwidth,trim=.25cm 0.25cm 0.25cm 0.25cm,clip]{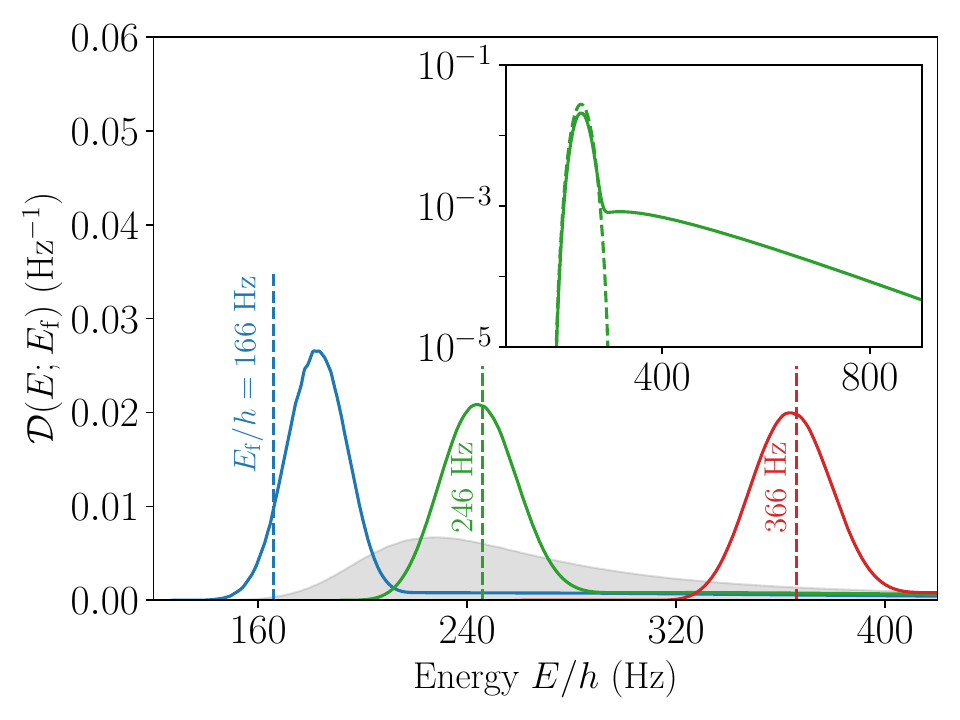}
\vspace*{-.25cm}
\caption{\justifying Estimated energy distribution $\mathcal{D}(E;E_{\rm f})$ of the atoms experimentally transferred into a random potential of amplitude $\VR/h=\SI{416}{\hertz}$, for three target energies $\Ef/h=166$ (left), $246$ (middle), and $366~\mathrm{Hz}$ (right), indicated by vertical dashed lines. The distributions $\mathcal{D}(E;E_{\rm f})$ are calculated using Eq.~(\ref{eq:Dtotal}), assuming a thermal fraction of 25$\%$ in state $|1\rangle$. Each distribution consists of a dominant BEC component, numerically computed from Eq.~(\ref{eq:Dtotal}), superimposed on a small, smooth thermal background [Eq.~(\ref{eq:Dthermal})], which exhibits a long tail toward higher energies. The inset displays the distribution for $E_{\rm f}/h = 246~\mathrm{Hz}$ on a semi-logarithmic scale, highlighting the high-energy thermal tail; the dashed line shows the result obtained for a pure BEC in state $|1\rangle$ ($\fc=1$). In the main panel, the shaded area shows the spectral function $A(E,\bk=0)$, whose numerical computation is detailed in Appendix~\ref{App:spectral}.}
\label{fig:EnergyDistribution}
\end{figure}

The thermal contribution describes the rf transfer into the disorder of the thermal atoms initially present in state $|1\rangle$. As shown in Appendix~\ref{App:energy_distr}, it generally involves a convolution of the energy filter $F(E)$ with the Boltzmann factor $\exp(-E/k_\mathrm{B}T)$. However, owing to the relatively large thermal scale $k_\mathrm{B}T/h \sim \SI{140}{\hertz}$ at $T=\SI{7}{\nano\kelvin}$, the resulting distribution $\mathcal{D}_{\rm th}(E;\Ef)$ essentially reduces, up to a prefactor given by the density of states $\rho(E)\propto\sqrt{E}$, to the Boltzmann weight $\exp[-(E-E_{\rm f})/k_\mathrm{B}T]$.

%The second term $\mathcal{D}_{\rm th}(E;E_{\rm f})$, accounts for the rf transfer of thermal atoms with a broad initial momentum distribution. In the narrow-filter limit, where the rf-filter $F(E-E_{\rm f})$ is much narrower than the spectral features, the rf pulse selects a narrow energy window around $E_{\rm f}$, over which the spectral function varies slowly. The thermal contribution then becomes insensitive to the detailed spectral shape and reduces to the Maxwell–Boltzmann distribution of the initial state at a fixed energy offset. This yields a square-root onset at $E_{\rm f}$, set by the density of states $\rho(E-E_{\rm f})\propto\sqrt{E-E_{\rm f}}$ and the Heaviside function $\Theta$, followed by an exponential decay governed by the temperature $1/k_\mathrm{B}T$.

Figure~\ref{fig:EnergyDistribution} shows energy distributions $\mathcal{D}(E;E_{\rm f})$ calculated numerically from Eq.~\eqref{eq:Dtotal} for three target energies, $E_{\rm f}/h = 166$, $246$ and $366~\mathrm{Hz}$ (indicated by vertical dashed lines), for a disorder amplitude of $V_{\rm R}/h = 416~\mathrm{Hz}$ and a thermal fraction of $1-\fc = 0.25$.  The shaded area represents the zero-momentum spectral function $A(E,\bk=0)$ that is used to calculate the energy distribution from Eq.~\eqref{eq:energy_dist} (see Appendix~\ref{App:spectral} for details on the numerical computation of the spectral function). Although the thermal contribution yields a broad pedestal (see inset of Fig.~\ref{fig:EnergyDistribution}), it remains small enough such that the energy distribution remains essentially very narrow, one~\cite{Jendrzejewski2012,Semeghini2015} or even two~\cite{kondov2011three, Barbosa2024} orders of magnitude smaller than in previous experiments.

%We characterize it by the HWHM width, which is approximately $\Delta E = 17~\mathrm{Hz} \ll \Ef$. \textcolor{blue}{[VJ: I beleive this last sentence is useless]}

%In contrast, within the rf-transfer scheme the energy distribution is much narrower and centered around $E_{\rm f}$, although it still exhibits a small but broad thermal pedestal. This feature is highlighted in the inset for $E_{\rm f}/h = 246~\mathrm{Hz}$, shown on a semi-logarithmic scale.
%Despite the presence of this shallow thermal background, a narrow energy resolution is achieved for the majority of the transferred atoms. We characterize it by HWHM, which is approximately $\Delta E = 17~\mathrm{Hz} \ll \Ef$.

\subsection{Comparison with 3D self-consistent theory}
\label{Sec:comparison_exp_sct}
\vspace{0.3cm}

We now compare the experimental density profiles after evolution in the speckle disorder with the SCT predictions. For this comparison, the SCT is supplied with three quantities: the initial experimental density profiles (Fig.~\ref{fig:Initialprofile_NJP}), the energy distribution $\mathcal{D}(E;\Ef)$ computed numerically from Eqs.~\eqref{eq:Dtotal} and (\ref{eq:Dthermal}), and the experimental value of the mobility edge, $\Ec^\text{exp}/h \simeq 237~\text{Hz}$. The latter was measured independently in Ref.~\cite{Josse2026} through a study of the inverse central-density dynamics, $n_\text{1d}^{-1}(z=0,t)$, thereby providing an unambiguous identification of the localized and diffusive phases.
Within our theoretical framework, the 1D integrated density profile after an evolution time $t$ follows from Eq.~\eqref{eq:density_full} as
\begin{align}
\label{eq:n1D_full}
    n_\text{1d}(z,t) = \int^\infty_{-\infty} \dd z^\prime
     \int^\infty_0&  \dd E\, \mathcal{D}(E;\Ef) P_{E}(z,z^\prime,t)n_\text{1d}(z^\prime,t=0) \;,  
\end{align}
where $P_{E}$ is the disorder-averaged propagator defined in Eq.~(\ref{eq:P}). To compute this integral, we use for the initial distribution $n_{\rm 1d}(z^\prime,t=0)$ fits of  the experimental density profiles measured immediately after the rf transfer to an asymmetric pseudo-Voigt  function (see Ref.~\cite{C8AN00710A}). Examples of these fits are shown as solid lines in Fig.~\ref{fig:Initialprofile_NJP}.

\begin{figure}[t]
\centering
\includegraphics[width=0.65\textwidth,trim=.25cm 0.25cm 0.25cm 0.25cm,clip]{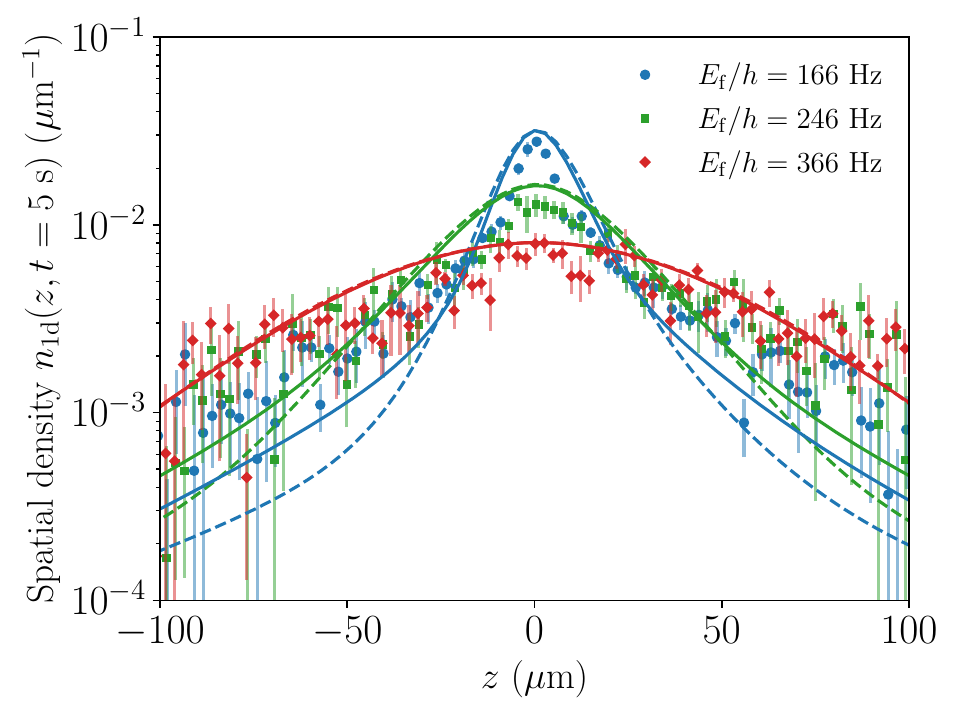}
\vspace*{-.25cm}
 \caption{\justifying 1D density profiles $n_{\rm 1d}(z,t)$, after an expansion of $t=\SI{5}{\second}$ in the disorder, for three energies: $\Ef/h = 166$, 246, and 366~Hz. Symbols are the experimentally measured profiles, while solid lines show the corresponding SCT predictions $\alpha_0 = 4.41 \pm 0.64~\mathrm{s}^{-1/3}$, $\beta = 2.84 \pm 0.34~\mu\mathrm{m}~\mathrm{s}^{-1/3}$. The pronounced differences between the profiles illustrate the Anderson transition as $\Ef$ crosses the mobility edge ($\Ec^\text{exp}/h \simeq 237$~Hz from Ref.~\cite{Josse2026}).
 The dashed curves show for comparison the prediction of the SCT assuming a pure BEC in state $|1\rangle$ ($\fc=1$ instead of $\fc=0.75$), demonstrating the importance of properly accounting for the  atomic energy distribution given by Eq.~\eqref{eq:Dtotal}. % \textcolor{blue}{[VJ: give values of $\alpha_0$ and $\beta$ for each case.]}
 }
\label{fig:Dprofile_fixedt}
\end{figure}
As explained in Sec.~\ref{Sec:Propagator}, the self-consistent theory depends on three effective parameters: the mobility edge $\Ec$, and the coefficients $\alpha_0$ and $\beta$. Here the mobility edge is fixed independently using the experimental determination $\Ec^\text{exp}/h\simeq 237\,\text{Hz}$ reported in Ref.~\cite{Josse2026} (and in agreement with transfer-matrix computations \cite{Pasek2017}). The parameters $\alpha_0$ and $\beta$ are calibrated according to the same method explained in Sec.~\ref{Sec:comparison_num_sct}, where the sum in Eq.~\eqref{Eq:optimization} runs over target energies spanning the mobility edge and over evolution times up to three seconds. This procedure is applied both to the ideal energy distribution $\mathcal{D}_\text{BEC}(E;\Ef)$—corresponding to a pure BEC in state $|1\rangle$—and to the full distribution $\mathcal{D}(E;\Ef)$ that accounts for the 25\% thermal fraction [see Eq.~\eqref{eq:Dtotal}].
\newline

We first show in Fig.~\ref{fig:Dprofile_fixedt} the experimental density profiles $n_{\rm 1d}(z,t)$ of atoms in state $|2\rangle$ along the $z$-axis at the maximum evolution time $t = \SI{5}{\second}$, for the three energies $\Ef/h=166$, $246$ and $366$~Hz considered so far. The pronounced change in the profile shapes illustrates the expected behavior across the mobility edge $\Ec^\text{exp} \simeq \SI{237}{\hertz}$: from a sharply peaked profile with exponentially decaying tails in the localized regime, to a broad profile in the diffusive regime.
The experimental data are compared with the theoretical prediction (\ref{eq:n1D_full}) using  either $\mathcal{D}_\text{BEC}(E;\Ef)$  ($\fc =1$, dashed curves) or $\mathcal{D}(E;\Ef)$ (with $\fc =0.75$, solid curves) for the energy distribution. In the diffusive regime, both choices provide a good description of the data. However, at the mobility edge and in the localized regime, significantly better agreement---particularly in the tails of the density profiles---is obtained when the thermal fraction is properly taken into account. This confirms the crucial role of the atomic energy distribution and highlights the need for an accurate characterization of the initial cloud in order to reproduce the observed spatial profiles (see also Refs.~\cite{Jendrzejewski2012,muller2014comment,Semeghini2015,Barbosa2024}). 

We note that even when thermal atoms are included, a residual discrepancy remains in the density tails in the localized regime. This may originate from effects present in the experiment but not included in the theory, such as interactions during the rf transfer or the residual shallow harmonic confinement in the $x$–$y$ plane. These mechanisms may also account for the values $\alpha_0 = 4.41 \pm 0.64~\mathrm{s}^{-1/3}$ and $\beta = 2.84\pm 0.34~\mu\mathrm{m}~\mathrm{s}^{-1/3}$  , which are slightly smaller, though of the same order of magnitude, than those obtained under ideal conditions in the numerical simulations of Sec.~\ref{Sec:numerics}.
\newline

\begin{figure}[t]
\centering
\includegraphics[width=\textwidth,trim=.25cm 0.3cm 0.25cm 0.25cm,clip]{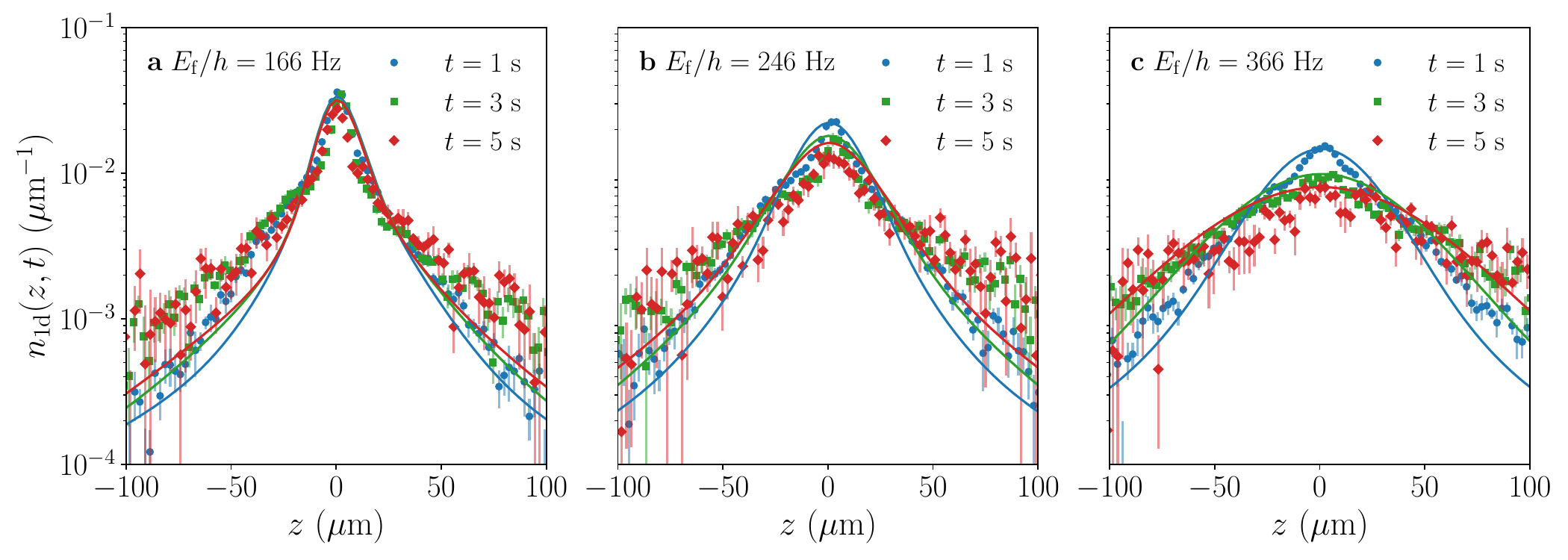} 
\vspace*{-.5cm}
\caption{\justifying Time evolution of the 1D density profiles $n_{\rm 1d}(z,t)$ in the disorder amplitude $\VR/h=\SI{416}{\hertz}$ for three target energies: (a) $\Ef/h=166$, (b) $246$, and (c) $366$~Hz. Symbols denote experimental data at $t=1$, 3, and $\SI{5}{\second}$ (semi-log scale). Solid lines represent the results of the SCT taking into account the thermal fraction $1-\fc=0.25$ in the energy distribution. The evolution of the profiles with energy and time reflects the transition from the diffusive to the localized regime. Each point is averaged over 6 to 9 independent experimental runs, and error bars indicate the standard error of the mean.}
\label{fig:Dprofile_vstime}
\end{figure}

We further show in Fig.~\ref{fig:Dprofile_vstime} the time evolution of 1D density profiles $n_{\rm 1d}(z,t)$ for the same three energies (points, experimental data), together with the prediction of the SCT taking into account the finite thermal fraction of the cloud (solid curves). The plots clearly reveal atomic spreading in the diffusive regime, while the profiles remain nearly time-independent in the localized regime. Figure~\ref{fig:Dprofile_vstime}a also suggests that, in the localized regime, the core of the distribution  appears slightly narrower than the initial density profile shown in Fig.~\ref{fig:Initialprofile_NJP}. 
We tentatively attribute this feature to the intrinsically strong spatial fluctuations of the localized eigenstates onto which the density profile decomposes. In the experiment, the averaging over disorder realizations is only partial, so that residual signatures of these fluctuations may persist in the measured profiles. Such effect lies beyond the scope of the SCT, which essentially provides a mean-field description of the Anderson transition, but would be worthwhile to investigate in future work.

\section{Conclusion}
\label{Sec:conclusion}

In this work, we have developed a flexible theoretical framework for quantitatively describing the dynamics of matter-wave packets in three-dimensional random potentials. The approach combines an accurate characterization of the atomic energy distribution with a three-dimensional implementation of the self-consistent theory of localization, providing access to the dynamics across the mobility edge.
Although the SCT is intrinsically approximate, we have shown that it accurately reproduces exact 3D numerical simulations in a nontrivial anisotropic speckle potential, while being significantly less computationally demanding. Within this framework, we have further achieved quantitative agreement with the density profiles reported in the recent experiment of Ref.~\cite{Josse2026}, where the dynamics near the Anderson transition could be probed using well-controlled atomic energy distributions.

Beyond the present study, our methodology can be readily extended to other platforms, such as fermionic systems~\cite{kondov2011three, Barbosa2024}. It would also be interesting to generalize the approach to more complex scenarios, for instance by including a shallow transverse confinement, as in the experiment of Ref.~\cite{Josse2026}, or by incorporating weak atom–atom interactions within the SCT, following the lines of previous works~\cite{Cherroret2016, Cherroret2014}.

On the experimental side, it would be highly appealing to apply the rf-transfer scheme to other physical problems. Relevant examples include measurements of critical exponents or multifractality~\cite{Akridas2019, Mirlin2000, mirlin2000b}, investigations of percolation transitions in random potentials~\cite{Pezze2011} and their connection to Anderson physics~\cite{Vrech2026}, exploration of localization phenomena in momentum space~\cite{CFS_Karpiuk2012, Jendrzejewski2012CBS, Arabahmadi2024}, and characterization of mobility edges and critical phases in one-dimensional quasi-periodic potentials~\cite{Biddle2010, Luschen2018, An2021MobilityEdge, Wang2022}.

\appendix

\section{Disorder-average density}
\label{App:theory}

In this appendix, we provide details on the derivation of the disorder-average density, Eq. (\ref{eq:density_full}).  Our starting point is the general relation (\ref{eq:density_WP}), in which we express the Green's function correlator in the hydrodynamic regime as \cite{Akkermans07}
\begin{align}
\overline{G^R(\br,\br',E_+)G^{A}(\br'',\br,E_-)}\simeq \frac{1}{2\pi\rho\tau^2}
\int \dd \br_1 \dd \br_2&|\overline{G}^R(\br,\br_2,E)|^2P_E(\br_1,\br_2,\omega)\nonumber\\
&\times\overline{G}^R(\br_1,\br',E)\overline{G}^{R*}(\br_1,\br'',E).
\end{align}
Here $\overline{G^R}$ ($\overline{G^A}$) is the averaged retarded (advanced) Green's function, $P_E(\br_1,\br_2,\omega)$ is the quantum propagator, which gives the probability density for an atom of energy $E$ to propagate from  $\br_1$ to $\br_2$, $\tau$ is the scattering mean free path and $\rho$ the density  of states per unit volume. Next, we insert this decomposition into Eq. (\ref{eq:density_WP}) and introduce the Wigner change of variables $(\br',\br'')\to[\boldsymbol{R}=(\br'+\br'')/2,\Delta\br=\br'-\br'']$. This leads to
\begin{align}
\label{eqdensity_general}
\overline{|\psi(\br,t)|^2}&= \int\frac{\dd E}{2\pi}\frac{\dd \omega}{2\pi}e^{-i\omega t}\!\!\int \dd \boldsymbol{R}\,\dd \Delta\br \dd \br_1 \dd\br_2
\overline{G}^R(\br_1,\boldsymbol{R}\!+\!\Delta\br/2,E)\overline{G}^{R*}(\br_1,\boldsymbol{R}\!-\!\Delta\br/2,E)\nonumber\\
&\times\frac{1}{2\pi\rho\tau^2} P_E(\br_1,\br_2,\omega)|\overline{G}^R(\br,\br_2,E)|^2
|f(E-\Ef)|^2
\phi(\boldsymbol{R}+\Delta\br/2)\phi^*(\boldsymbol{R}-\Delta\br/2).
\end{align}
Here, we have  further approximated $f(E_+-\Ef) f^*(E_- - \Ef)\simeq |f(E-\Ef)|^2$, using that in the long-time dynamics, $t\gg\tau$, the dominant contribution to the integral over $\omega$ arises from the vicinity of $\omega= 0$ owing to the rapidly oscillating factor $\exp(-i \omega t)$, so that $E_\pm = E \pm \frac{\omega}2 \sim E$. 
To simplify this expression, we proceed in two steps. First, instead of directly working with the product $\phi\phi^*$ of initial wave functions in real space, it is convenient to introduce the Wigner distribution $W(\boldsymbol{R},\bk)$ of the initial state, defined by the relation:
\begin{equation}
\phi(\boldsymbol{R}+\Delta\br/2)\phi^*(\boldsymbol{R}-\Delta\br/2)=
\int\frac{\dd \bk}{(2\pi)^3}e^{i\bk\cdot\Delta\br}W(\boldsymbol{R},\bk).
\end{equation}
The main interest of the Wigner distribution is to simultaneously encode the spatial \emph{and} momentum structures of the initial state. In particular, it gives access to the spatial density after integration over the momentum variable:
\begin{equation}
\label{eq:Wformula}
\int\frac{\dd\bk}{(2\pi)^3} W(\boldsymbol{R},\bk)=\overline{|\phi(\br)|^2}.
\end{equation}
Second, we use that at scales larger than the mean free path (hydrodynamic regime), $P_E(\br_1,\br_2,\omega)$ is a smooth function of $\br_1$ and $\br_2$, so that $P_E(\br_1,\br_2,\omega)\simeq P_E(\boldsymbol{R},\br,\omega)$. With this simplification, the three integrals over $\Delta\br$, $\br_1$ and $\br_2$ in Eq. (\ref{eqdensity_general}) can be easily performed,
%\footnote{We use, in particular, that $\int d\br_2 |\overline{G}^R(\br,\br_2,E)|^2=2\pi\rho\tau$.}
yielding:
 \begin{align}
 \label{eq:psi2_appendix}
\overline{|\psi(\br,t)|^2}\simeq \frac{1}{\tau} \int\frac{\dd E}{2\pi}\!\!\int \dd \boldsymbol{R}\int\frac{\dd \bk}{(2\pi)^3}
|\overline{G}^R(\bk,E)|^2 |f(E-\Ef)|^2P_E(\boldsymbol{R},\br,t)W(\boldsymbol{R},\bk).
\end{align}
In the last step, we note that the momentum width of $|\overline{G}^R(\bk,E)|^2$ is of order $1/\ell$, where $\ell$ is the scattering mean free path, whereas the Wigner distribution has a momentum width $\sim 1/R_0$, with $R_0$ the spatial extent of the initial wave packet. In the experiment \cite{Josse2026}, one typically has  $R_0\gg\ell$, so that in Eq.~(\ref{eq:psi2_appendix}) one can approximate $\smash{|\overline{G}^R(\bk,E)|^2 \simeq |\overline{G}^R(\bk=0,E)|^2}$.
This enables us to perform the integral over $\bk$ using Eq.~(\ref{eq:Wformula}). Noting further that $\smash{|\overline{G}^R(0,E)|^2 \equiv 2\pi \tau A(E,\bk=0)}$, where $A$ is the spectral function, we recover Eq.~(\ref{eq:density_full}) of the main text.

\section{Energy distribution of the atomic cloud}
\label{App:energy_distr}

In this appendix, we present a derivation of the energy distribution (\ref{eq:Dtotal}), which accounts for a finite thermal contribution given by Eq. (\ref{eq:Dthermal}). The starting point is the density matrix of the Bose gas  in state $|1\rangle$ prior to transfer:
\begin{align}
\label{eq:rho}
\hat{\rho}= \fc |\bk=0\rangle \langle\bk=0|+(1-\fc)\frac{e^{-\hat{H}_0/k_BT}}{Z},
\end{align}
where $\hat{H}_0$ is the free-space Hamiltonian, $Z$ a normalization factor, and we have approximated  $|\phi\rangle\simeq |\bk=0\rangle$ for the BEC wave function. The rf pulse transfers an initial energy $E_1$ in state $|1\rangle$ to a final energy $E_2=E_1+\hbar\omega_\text{rf}$ in the disorder-sensitive state $|2\rangle$. 
For an initial $|\bk\rangle$ state, $E_1=E_k\equiv \hbar^2\bk^2/(2m)$, and the transfer  in the disorder corresponds to the transformation
$|\bk\rangle\to f(\hat{H}-E_k-\Ef)|\bk\rangle$, which generalizes Eq. (\ref{eq:expansion_psi0}), with $\Ef\equiv \hbar\omega_\text{rf}$ the transferred energy. The energy distribution in the disorder after transfer is defined as 
\begin{align}
\mathcal{D}(E)\equiv \text{Tr}\Big[\overline{\hat{\rho}\,\delta(E-\hat{H})}\Big].
\end{align}
The initial BEC state $|\bk=0\rangle$ has energy $E_1=0$, so its energy distribution after transfer in the disorder reads
\begin{align}
\label{eq:ABEC}
\mathcal{D}_\text{BEC}(E)=
 \fc\langle\bk=0| f^*(\hat{H}-\Ef) \overline{\delta (E-\hat{H})}f(\hat{H}-\Ef)|\bk=0\rangle,
\end{align}
which immediately leads to Eq. (\ref{eq:energy_dist}) of the main text.

Consider now the energy distribution of the thermal component. Since the thermal factor $\smash{e^{-\hat{H}_0/k_BT}}$ in Eq. (\ref{eq:rho}) involves the free Hamiltonian, it is convenient to introduce the basis states $\{E_k,|\bk\rangle\}$ to evaluate the trace: 
\begin{align}
\mathcal{D}_\text{Th}(E)&=\frac{1-\fc}{Z}\int \frac{\dd\bk}{(2\pi)^3}\overline{ \langle \bk|f^*(\hat{H}-E_k-\Ef) e^{-\hat{H}_0/k_BT}\delta(E-\hat{H})f(\hat{H}-E_k-\Ef)| \bk\rangle}\nonumber\\
&=
\frac{1-\fc}{Z}\int \frac{\dd \bk}{(2\pi)^3}
F(E-E_k-\Ef)e^{-E_k/k_BT} A(E,\bk).
\end{align}
We then approximate $ A(E,\bk)\simeq A(E-E_k,\bk=0)$, and change variable in the momentum integral from $\bk$ to $E'=\hbar^2\bk^2/(2m)$. This leads to:
\begin{align}
\mathcal{D}_\text{Th}(E)=\frac{1-\fc}{Z}\int \dd E'\rho(E')
F(E-E'-\Ef)e^{- E'/k_BT} A(E-E',\bk=0),
\end{align}
with $\rho$ the free-space density of states in three dimensions. We finally perform the integral over $E'$ using that the filter function is much narrower than the spectral function, the thermal factor and the density of states:
\begin{align}
\mathcal{D}_\text{Th}(E)\simeq\frac{1-\fc}{Z}\rho(E-\Ef)A(\Ef,\bk=0)\,
e^{- (E-\Ef)/k_BT}.
\end{align}
Using that $\rho(x)=\Theta(x)\sqrt{x}$ in three dimensions (with $\Theta$ the Heaviside step function), and imposing the normalizing condition $\int dE\, \mathcal{D}_\text{th}(E)=1-\fc$, we finally obtain Eqs. (\ref{eq:Dthermal}) of the main text.

\section{Numerical methods}
\label{App:numerical_methods}
\vspace{0.3cm}

\subsection{Numerical generation of the speckle potential}
\label{App:speckle}
Experimentally, the speckle potential is generated by illuminating a diffusive lens with a laser beam of vacuum wavelength $\lambda = 780$~nm (wavenumber $k_L = 2\pi/\lambda = 8.06$~\textmu$\mathrm{m}^{-1}$). The incident beam is Gaussian, with a waist of $w= 9.9$~mm at the lens plane, and is truncated by a diaphragm of diameter $D = 20.4$~mm. The lens has a focal length $d = 15.2$~mm. The beam propagates along the $x$-axis and is linearly polarized in the $(y,z)$-plane.
Numerically, the speckle potential is computed using a Fourier-filtering method similar to that in Ref.~\cite{Volchkov2018}. The electric field along the polarization direction is expressed in the Fourier representation as
\begin{equation}
    E (\br) = E_0 \int_{k_x>0} \frac{\dd \bk}{(2\pi)^3} \xi (\bk) \frac{k_x}{k_L}\exp \left( - \frac{k_\parallel^2 d^2}{k_L^2 w^2} \right) \Theta \left(k_c - k_\parallel \right) \delta (|\bk| - k_L) \exp \left( i \bk \cdot \br \right) \: \: .
    \label{eq:Fourier:expansion}
\end{equation}
Here, $E_0$ is an arbitrary amplitude, and $\xi$ is a uncorrelated, complex Gaussian stationary random process with zero mean. The wave vector $\bk_\parallel = (k_y,k_z)$ denotes the projection of $\bk$ onto the $(y,z)$-plane, and $\smash{k_c = k_L [(D/2d)^2/(1 + (D/2d)^2)]^{1/2}}$ is the cutoff set by the diaphragm. The Dirac distribution $\delta (|\bk| - k_L)$ enforces the light’s dispersion relation in vacuum. Numerically, the integral~\eqref{eq:Fourier:expansion} is evaluated using a 3D fast Fourier transform on a rectangular grid, approximating  the Dirac distribution by a narrow Gaussian of width $\Delta k$:  $\delta (|\bk| - k_L) \approx \exp\left[ - (|\bk| - k_L)^2/(2\Delta k^2) \right] / (\sqrt{2 \pi} \Delta k)$.\medskip

Equation~\eqref{eq:Fourier:expansion} implies that the field is a sum of independent Gaussian random variables $\xi(\bk)$. Consequently, the probability density of the potential $V(\br) = K |E(\br)|^2$, where $K$ is a constant characterizing the light–atom interaction, follows the exponential distribution $P(V) = \exp( -V/\VR)/\VR \cdot \Theta(V)$~\cite{Goodman2007}. The average potential $\VR$ is obtained from $\VR = K \overline{|E(\br)|^2}$, using the uncorrelated nature of $\xi$: $\smash{\overline{\xi(\bk)\xi^*(\bk^\prime)} = (2 \pi)^3 \delta(\bk - \bk^\prime)}$. This yields $\VR=K |E_0|^2 \mathcal{I}$, where
\begin{equation}
	\mathcal{I} = \int \frac{\dd \bk }{(2\pi)^3} \frac{k_x^2}{k_L^2}\exp\left( - \frac{2 k_\parallel^2 d^2}{k_L^2 w^2} \right) \Theta\left(k_{c} - k_\parallel \right) \delta(|\bk| - k_L).
	\label{eq:VR}
\end{equation}
To numerically generate the speckle potential shown in Fig.~\ref{Fig:speckle}a–b, we proceed as follows. We evaluate Eq.~(\ref{eq:Fourier:expansion}) with $E_0 = 1$ using a numerically generated random noise $\xi$ (the corresponding numerical parameters are given in Sec.~\ref{Sec:Speckle potential}). The resulting field is normalized by $\mathcal{I}$, and the potential is then obtained as $V(\br) = \VR |E(\br)|^2/\mathcal{I}$, where $\VR=416\,\text{Hz}$ is the experimental rms disorder strength.

From the numerically generated speckle potential, the autocorrelation function shown in Fig.~\ref{Fig:speckle}c is computed as $C (\Delta\br) =  \overline{V(\br) V(\br+\Delta\br)}/\VR^2-1$. In the same figure, this result is compared with the theoretical prediction, which follows directly from Eq.~(\ref{eq:Fourier:expansion}) using the Gaussian character of $\xi$:
\begin{equation}
    C (\Delta \br) = \frac{1}{\mathcal{I}^2}\displaystyle\left| \int \frac{\dd \bk }{(2\pi)^3} \frac{k_x^2}{k_L^2} \exp\left( - \frac{2 k_\parallel^2 d^2}{k_L^2 w^2} \right)  \Theta\left(k_{c} - k_\parallel \right) \delta(|\bk| - k_L) \,  \exp\left( i \bk \cdot \Delta\br\right) \right|^2 \: .
    \label{eq:autocorr}
\end{equation}
%Note that, according to Eq.~\eqref{eq:VR}, explicit values of $K$ and $E_0$ are not required to generate the speckle potential. Numerically, we simply set $E_0 = 1$ in Eq.~\eqref{eq:Fourier:expansion} and $K = 1$ in arbitrary units. The properly scaled potential is then obtained as $V(\br) = \VR |E(\br)|^2/\mathcal{I}$. \textcolor{red}{[NC: this is not clear]} {\color{blue}JP nouvelle suggestion : Note that, according to Eq.~\eqref{eq:VR}, explicit values of $K$ and $E_0$ are not required to generate the speckle potential with an appropriately scaled amplitude. Indeed, it suffices to provide the product $K |E_0|^2 = \VR/\mathcal{I}$.} \medskip

%Figure~\ref{Fig:speckle} of the main text shows a realization of the speckle potential computed numerically from Eq. (\ref{})
%In obtaining these results, we have used a cubic domain size of length $L=100$~\textmu m,  discretization steps $\Delta x = 0.7$~\textmu m and  $\Delta y = \Delta z = 0.3$~\textmu m,  and $\Delta k = k_L/100$. 
%The dots corresponding to the numerical data allow one to appreciate that a sufficient number of points indeed resolve the speckle grain along each direction (about 5 points along both the $y$ and $z$-directions and about 11 points along the $x$-direction). 

\subsection{Chebyshev polynomial expansion of operators}
\label{App:chebyshev}

The numerical computation of the initial energy-resolved state, the spectral function, and the time evolution of the wave packet presented in the main text rely on a unified framework based on Chebyshev polynomial expansions of operator functions of the Hamiltonian. The calculations are performed on the spatial grid used to generate the speckle potential, with the Laplacian term of the Hamiltonian discretized using a three-point finite-difference stencil in each spatial direction.
%For these three problems, the wave function $\psi$ and the Hamiltonian $\hat H$ are discretized on the spatial grid used to generate the speckle potential, employing a three-point finite-difference stencil in each direction for the Laplacian.

In the three cases, the problem reduces to applying a given operator function of the Hamiltonian, $\mathscr{F}(\hat H)$, to an initial state.
%Common to all tasks is the discretization of the wave function $\psi$ and of the Hamiltonian $\hat{H}$ by finite differences on the spatial grid on which we defined the speckle with a three-point finite difference stencil along each direction for the Laplace operator.  For all tasks mentioned above, we will see that they amount to applying a certain function of the Hamiltonian $\mathscr{F}(\hat{H})$ on some state. 
Since full diagonalization of the 3D Hamiltonian would be computationally prohibitive, we approximate $\mathscr{F}(\hat H)$ by a truncated Chebyshev expansion. To this end, the discrete Hamiltonian is first shifted and rescaled so that its spectrum lies in the interval $(-1,1)$, where Chebyshev polynomials are defined. Given bounds $E_{\min}$ and $E_{\max}$ of the spectrum, we define $E_\mathrm{mid} = (E_{\max} + E_{\min}) / 2$ and introduce the rescaled operator $\tilde{H} = 2 (\hat{H} - E_\mathrm{mid})/(E_{\max} - E_{\min} )$, whose eigenvalues lie in $(-1,1)$.
%Applying a function of an operator is easy provided we are given an eigen-basis of this operator.  However,  diagonalizing the Hamiltonian would be prohibitive for the large three-dimensional system studied in this work.  Instead, we approximate $\mathscr{F}(\hat{H})$ using a truncated expansion in a fast converging series of Chebyshev polynomials. To do so, we need first to rescale the discrete Hamiltonian in such a way as to map its eigenvalues into the interval $(-1,1)$ on which the Chebyshev polynomials are defined.  
%Given a lower bound $E_{\min}$ and an upper bound $E_{\max}$ for the discrete spectrum, we may construct the shifted and rescaled Hamiltonian $\tilde{H} = 2 (\hat{H} - E_\mathrm{mid})/(E_{\max} - E_{\min} )$ with $E_\mathrm{mid} = (E_{\max} + E_{\min}) / 2$. The eigen-energies $\varepsilon_n$ of $\tilde{H}$ are then of the form $\varepsilon_n = 2 (E_n - E_{\mathrm{mid}})/(E_{\max} - E_{\min})$ and lie in the interval $(-1,1)$. 
%Using the fundamental property of the Dirac distribution, we may write
Using
\begin{equation}
    \mathscr{F}(\tilde{H}) = \int_{-1}^{1} \delta (\varepsilon - \tilde{H}) \mathscr{F}(\varepsilon) \: \dd\varepsilon \: ,
    \label{eq:dirac:trick}
\end{equation}
 together with the Chebyshev expansion for the Dirac distribution~\cite{Weisse2006}, 
\begin{equation}
    \delta (\varepsilon - \tilde{H}) = \frac{1}{\pi \sqrt{1 - \varepsilon^2}} \left[ T_0 (\tilde{H}) + 2 \sum_{n=1}^\infty T_n (\tilde{H}) T_n (\varepsilon) \right] \: ,
    \label{eq:dirac_chebyshev}
\end{equation}
one obtains the Chebyshev expansion 
\begin{align}
\label{eq:chebyshev_expansion}
\mathscr{F}(\tilde{H}) = c_0 T_0(\tilde{H}) + 2 \sum_{n=1}^\infty c_n T_n(\tilde{H})\:, \ \ \ 
c_n = \int_{-1}^{1} \frac{\mathscr{F}(\varepsilon) T_n(\varepsilon)}{\pi \sqrt{1 - \varepsilon^2}} \: \dd\varepsilon \: ,
\end{align}
where $T_n$ denotes the Chebyshev polynomial of the first kind of order $n$. 
Applying $\mathscr{F}(\tilde H)$ to a state $\Ket{\phi}$ thus amounts to computing the vectors $\Ket{n}=T_n(\tilde H)\Ket{\phi}$, generated recursively via
%\begin{equation}
 %   \mathscr{F}(\tilde{H}) \Ket{\phi} = c_0 T_0(\tilde{H}) \Ket{\phi} + 2\sum_{n=1}^\infty c_n T_n (\tilde{H}) \Ket{\phi} = c_0 \Ket{0}  + 2 \sum_{n=1}^\infty c_n \Ket{n} \: ,
%    \label{eq:Fphi}
%\end{equation}
\begin{align}
\label{eq:recursion}
    \Ket{0} = \Ket{\phi},\ \ \ \ \Ket{1} = \tilde{H} \Ket{0},\ \ \ \ \Ket{n+1} = 2 \tilde{H} \Ket{n} - \Ket{n-1}.
\end{align}
%where we have defined the vectors $\Ket{n} \equiv T_n(\tilde{H}) \Ket{\phi}$.  They are obtained by the recurrence inherited from the Chebyshev polynomials as $\Ket{0} = \Ket{\phi}$, $\Ket{1} = \tilde{H} \Ket{0}$ and $\Ket{n+1} = 2 \tilde{H} \Ket{n} - \Ket{n-1}$.  
Each new vector requires only a sparse matrix–vector multiplication and a vector subtraction, making the method both efficient and numerically stable. In practice, the series is truncated at finite order to reach the desired accuracy and evaluated using the Clenshaw algorithm~\cite{Clenshaw1955}.
%Each new vector $\Ket{n+1}$ is thus obtained from the two previous ones at the low computational cost of a sparse matrix-vector multiplication and a vector subtraction.  Numerically the expansion is truncated to achieve a desired precision and summed with the help of the Clenshaw algorithm for improved efficiency and numerical stability~\cite{Clenshaw1955}.

\subsection{Numerical generation of the energy-resolved state}
\label{App:filter}
We first show how the Chebyshev method is used to numerically generate the energy-resolved state $\Ket{\psi_0}$, defined by Eq.~\eqref{eq:expansion_psi0}. In this problem, on applies the operator $\mathscr{F}(\hat{H}) = f(\hat{H} - \Ef)$ to the BEC state $\Ket{\phi}$ prior to transfer, where $f$ is the filter function discussed in the main text.  
In this case, we use $\Ket{\psi_0}=\mathscr{F}(\tilde{H}) \Ket{\phi} = c_0 \Ket{0}  + 2 \sum_{n=1}^\infty c_n \Ket{n}$, 
%The Chebyshev expansion presented above provides an efficient method for applying this operator on the state $\Ket{\phi}$. Indeed, it suffices to directly apply the Eq.~\eqref{eq:Fphi} 
where the coefficients $c_n$ 
%of Eq.~\eqref{eq:coeff} must be 
are computed from Eq. (\ref{eq:chebyshev_expansion}). % for the filter function $\tilde{f}$ in the rescaled energy variable $\varepsilon$ defined by $\tilde{\mathscr{F}}(\varepsilon) = \mathscr{F}(E)$.  
The integral defining $c_n$ in Eq.~\eqref{eq:chebyshev_expansion} is conveniently evaluated by introducing the change of variable $\varepsilon = \cos\theta$. Using the identity $T_n(\varepsilon) = \cos(n\arccos \varepsilon)$, the coefficients can be written as the cosine transform
\begin{equation}
    c_n = \frac{1}{\pi} \int_0^\pi {\mathscr{F}}(\cos \theta) \cos (n \theta) \: \dd\theta \: .
\end{equation}
Numerically, the variable $\theta$ is sampled uniformly in the interval $(0,\pi)$ with $N$ points, $\theta_k = (2 k + 1) \pi / (2N) $ ($0 \leq k \leq N-1$), and the integral is approximated by the discrete cosine transform
\begin{equation}
    c_n \approx \frac{\Delta \theta}{\pi} \sum_{k=0}^{N-1} h_k \cos \left[n \frac{(2 k + 1) \pi}{2N} \right] \:, 
    %= \frac{\Delta \theta}{2 \pi} \mathrm{DCT} (\mathbf{h}) [n] \: ,
\end{equation}
where $\Delta\theta = \pi/N$ and $h_k = \mathscr{F}(\cos\theta_k)$.
%The last equality in the above equation is written in terms of the so-called discrete cosine transform, DCT of type II. 

\subsection{Numerical generation of the spectral function}
\label{App:spectral}

The computation of the energy distribution $\mathcal{D}_\text{BEC}=F(E-\Ef)A(E,\bk=0)$ appearing in Eq.~(\ref{eq:Dtotal}) requires the numerical evaluation of the spectral function $A(E,\bk=0)$. This is performed using a Chebyshev expansion with $\mathscr{F}(\hat H)=\delta(E-\hat H)$. To this end, we write:
\begin{equation}
A(E,\phi) \equiv \Bra{\phi} \delta \big( E - \hat{H} \big) \Ket{\phi} = \Bra{\phi} \delta \left[ \frac{E_{\max} - E_{\min}}{2}(\varepsilon - \tilde{H}) \right] \Ket{\phi} = \frac{2 \tilde{A}(\varepsilon,\phi)}{E_{\max} - E_{\min}}  \: ,
\end{equation}
where 
\begin{equation}
\tilde{A}(\varepsilon,\phi) = \Bra{\phi} \delta \big( \varepsilon - \tilde{H} \big) \Ket{\phi}
\end{equation}
is the spectral function for the rescaled Hamiltonian, and we have introduced $|\phi\rangle=|\bk=0\rangle$.
Using the Chebyshev expansion of the Dirac distribution [Eq.~\eqref{eq:dirac_chebyshev}], we obtain
\begin{equation}
\tilde{A}(\varepsilon, \phi) = c_0 (\varepsilon) \BraKet{\phi}{0} + 2  \sum_{n=1}^{\infty} c_n (\varepsilon) \BraKet{\phi}{n} \quad \text{with} \qquad
c_n(\varepsilon) = \frac{T_n(\varepsilon)}{\pi \sqrt{1-\varepsilon^2}} \: .
\end{equation}
To suppress Gibbs oscillations arising from truncation at finite order $N$, the coefficients are multiplied by Jackson weights. This procedure—equivalent to convolving the spectral function with an optimal smoothing kernel—is known as the Kernel Polynomial Method~\cite{Weisse2006}.
For the energy distributions shown in Fig.~\ref{fig:EnergyDistribution}, we use $N=10^4$, ensuring that the Jackson broadening remains well below the intrinsic width of the filtered spectral functions (below $1\%$ for the narrowest spectral function of 5~Hz).

\subsection{Time evolution}
\label{App:time_evolution}

The initial state is propagated in time with step $\Delta t$ according to $\Ket{\psi(t+\Delta t)} = \exp \big( - i \hat{H} \Delta t \big) \Ket{\psi(t)}$. Numerically, the evolution operator is applied efficiently using a Chebyshev expansion with $\mathscr{F}(\hat H)=\exp(-i\hat H \Delta t)$~\cite{Fehske2009}.
To this end, the evolution operator is first rewritten as
\begin{equation}
    \hat{U} = \exp\left( -i E_{\mathrm{mid}} \Delta t \right) \exp \left( -i \tilde{H} \Delta \tilde{t} \right) = \exp\left( -i E_{\mathrm{mid}} \Delta t \right) \tilde{U} \: ,
    \label{eq:evolution:rescaling}
\end{equation}
where $\Delta\tilde t = (E_{\max}-E_{\min})\Delta t/2$.
The first factor in Eq.~\eqref{eq:evolution:rescaling} is a global  phase, which leads to an irrelevant energy shift  and has no impact on the dynamics. The second factor, denoted $\tilde{U}$, can be expanded as 
\begin{equation}
    \tilde{U} = \int_{-1}^{1} \delta (\varepsilon - \tilde{H}) \exp \left( -i \varepsilon \Delta \tilde{t} \right) \: \dd\varepsilon = c_0(\Delta \tilde{t}) T_0(\tilde{H}) + 2 \sum_{n=1}^\infty c_n(\Delta \tilde{t}) T_n(\tilde{H}) \: ,
\end{equation}
where Eq. (\ref{eq:dirac_chebyshev}) has been used in the second equality.
%In the second equality, we have used the expansion of the Dirac distribution in the basis of Chebyshev polynomials $T_n$, and interchanged the discrete sum and the integral over the rescaled energy $\varepsilon$. 
The coefficients $c_n(\Delta \tilde{t})$ are given in closed form as
\begin{equation}
    c_n(\Delta\tilde{t}) = \int_{-1}^{1} \frac{\exp(-i \varepsilon \Delta\tilde{t})}{\pi \sqrt{1-\varepsilon^2}} T_n(\varepsilon) \: \mathrm{d}\varepsilon = (-i)^n J_n(\Delta \tilde{t}) \: ,
\end{equation}
where $J_n$ denotes the Bessel function of the first kind of order $n$. Applying the evolution operator thus amounts to computing
\begin{equation}
    \Ket{\psi(t+\Delta t)} = \exp\left( -i E_{\mathrm{mid}} \Delta t \right) \left[ c_0(\Delta \tilde{t}) \Ket{0} + 2 \sum_{n=1}^\infty c_n(\Delta \tilde{t}) \Ket{n} \right] \: ,
\end{equation}
where the vectors $\Ket{n} = T_n(\tilde{H}) \Ket{\psi(t)}$ are generated recursively using Eq. (\ref{eq:recursion}). Owing to the rapid decay of the Bessel functions with $n$, the series can be safely truncated without additional regularization.

\ack{
The authors thank V. Denechaud, B. Lecoutre, A. Signoles for early work and D. Clément, D. Delande and M. Filoche for fruitful discussions. }

\funding{This work benefited financial support from the project Localization of Waves of the Simons Foundation (Grant No. 601939), the French National Research Agency under Grant No. ANR-24-CE30-6695 (FUSIoN), Grant No. ANR-25-CPJ1-0059-01 (CPJ LUMIS), Grant No. ANR-22-CMAS-0001(QuanTEdu-France) and Region Île-de-France in the framework of DIM QuanTiP.}

\roles{
 J.-P. B.,  S. B. and N. C. conducted the numerical simulations and developed the theoretical predictions. K. X., H.-M. Q., X. Y., M. N. and Y. G. carried out the experiments. All authors contributed to data analysis, project development, and the writing of the paper.}

% List author names and the contributions made to the article, using terms from the NISO Contributor Roles Taxonomy (CRediT) https://credit.niso.org

\data{All numerical data presented in this work are available under reasonable request.}

%\suppdata{Sample text inserted for demonstration.}

%\bibliographystyle{iopart-num}
\bibliographystyle{unsrt}
\bibliography{NJP_bib_fixed}
\end{document}